\documentclass[a4paper,fleqn]{cas-sc}

\usepackage[numbers]{natbib}
\usepackage[]{algorithm2e}
\usepackage{graphicx}
\usepackage{subcaption}
\usepackage{hhline}
\usepackage[utf8]{inputenc}
\def\tsc#1{\csdef{#1}{\textsc{\lowercase{#1}}\xspace}}
\tsc{WGM}
\tsc{QE}
\tsc{EP}
\tsc{PMS}
\tsc{BEC}
\tsc{DE}
\DeclareUnicodeCharacter{200E}{}
\begin{document}
\shorttitle{Biased Random Walk (BRW)}
\shortauthors{Hanif Emamgholizadeh et~al.}

\title [mode = title]{A Framework for Quantifying Controversy of Social Network Debates Using Attributed Networks: Biased Random Walk (BRW)}                      



\author[1]{Hanif Emamgholizadeh}[]
\address[1]{h.emamgholizadeh@gmail.com}

\author[2]{Milad Nourizade}[]
\address[2]{milad.nouriezade@gmail.com}

\author[3]{Mir Saman Tajbakhsh}[]
\address[3]{s.tajbakhsh@it.uut.ac.ir}

\author[4]{Mahdieh Hashminezhad}[]
\address[4]{hasheminezhad@yazd.ac.ir}

\author[5]{Farzaneh Nasr Esfahani}[]
\address[5]{farzane.nasresfahani@gmail.com}
 
\begin{abstract}
All societies have been much more bipolar over the past few years, particularly after the emergence of online social networks and media. In fact, the gap between two ends of social spectrum is going to be even deeper after the spread of new media. In this circumstance, social polarization has been a growing concern among socialists and computer science experts because of the detrimental impact which online social networks can have on societies by adding fuel to the fire of extremism.\\
Several researches were conducted for proposing measures to calculate controversy level in social networks, afterward, to reduce controversy among contradicting viewpoints, for example, by exposing opinions of one side to other side's members. Most of the attempts for quantifying social networks' controversy have considered the networks in their most primary forms, without any attributes. Although these kinds of researches provide platform-free algorithms to be used in different social networks, they are not able to take into account a great deal of useful information provided by users (node attributes).To surmount this shortcoming, we propose a framework to be utilized in different networks with different attributes. We propeled some Biased Random Walks (BRW) to find their path from start point to an initially unknown end point with respect to initial energy of start node and energy loss of nodes on the path. We extracted structural attribute of networks, using node2vec, and compared it with state-of-the-art algorithms, and showed its accuracy. Then, we extracted some content attributes of user and analyze their effects on the results of our algorithm. BRW is compared with another state-of-the-art controversy measuring algorithm. Then, its changes in different level of controversy in Persian Twitter is considered to show how it works in different circumstance. 
\end{abstract}


\maketitle

\section{Introduction}
The surge of online social network and media in the first years of new century seemed to be so promising not only for personal relations but also for social and political freedom. Then, gradually the darker side emerged. Although in an individual level social network provides wider communication, users try to make relations with like-minded people \cite{barbera2015tweeting,quattrociocchi2016echo}. In social level bots, cyborgs and trolls attempt to manipulate public opinion and even determine elections' results \cite{allcott2017social}. One of the growing concerns in this regard is the users' behaviors in controversial topics. Usually, controversial topics separate each community into, at least, two well-separated sub-communities with disparate opinions about the current issues, and online social networks escalate the controversy level by \textit{filter bubble} and also \textit{users selective exposure}, in which different sides of a debate are exposed only to the contents produced by other users with similar ideas \cite{messing2014selective}. This can make society even further radical and bipolar.\\

Quantifying controversy was the objective of social researches for the past decades \cite{isenberg1986group,dimaggio1996have,fiorina2008political,baldassarri2008partisans}. To do so, sociologists considered qualitative data and defined polarization as people's distance from ideological perspective \cite{baldassarri2008partisans}. They conducted survey-based research for evaluating participants' opinions about controversial topics. Afterward, conclusion were drawn by analyzing the results. The advent of social media in new century and huge free-to-use information provided by networks and users pushed sociology and computer science toward finding new methods for evaluating polarization in societies. Adamic and Glance's research \cite{adamic2005political} was the first study which made a bridge between offline polarization research and online world, their results were supported by Hargittai et al.s' qualitative and quantitative analyses \cite{hargittai2008cross}. These two researches concentrated on weblog's links for assessing users' behavior regarding controversial issues. Conover et al. \cite{conover2011political} extended previous research to social media. Recent researches also have confirmed the polarization, especially in political discussions \cite{alves2019together}.\\

After proving the existence of polarization, the next step is evaluating the amount of topic controversy which divides the society into two bipolar communities. A variety of measures from different perspectives have been proposed. On one hand, sociologists have been trying to find the distance of rival opinions \cite{bramson2016disambiguation}. On the other hand, computer scientists have been working with quantitative data in graphs or provided content in social networks \cite{garimella2018quantifying,morales2015measuring}.\\

Assessing and quantifying controversy is practical for detecting and predicting controversial topics \cite{popescu2010detecting}, reducing polarization \cite{garimella2017reducing}, and also finding fake news \cite{vicario2019polarization}. Additionally, for better prediction of information diffusion \cite{hoang2018predicting} controversial topics should be considered deeply. It seems that polarization and controversy detection is going to be the focus of more research in near future.\\

The past researches about detecting controversial topic contained some shortcomings. For example, some of them considered only a small part of a network, which are boundary nodes \cite{guerra2013measure}; others only rely on network structure \cite{garimella2018quantifying}. Finally, some of the algorithms are useful for special networks, such as signed networks \cite{bonchi2019discovering}. However, all of these measures have a common drawback; all of them ignore invaluable content and profile information provided by users in social networks. \\

To overcome this main drawback, we propose a framework for using in different platforms and networks with different kinds of attributes. Our framework is based on a biased random walk (BRW) in which there is an initial energy to start a journey by random walker, and in each step a special amount of energy, called as energy loss, is reduced by the node through which random walker passes. Based on the network and expected usage, there are several methods for assigning initial and energy loss; however, we have some suggestions for initial and energy loss, which our assessments demonstrate to be useful for measuring controversy level of a topic.\\

The rest of this paper is organized as follows. In the section \ref{sec_related}, we provide an extensive review on the proposed measures for quantifying controversy. In section \ref{sec_method}, we introduce our framework and suggest some variations for calculating controversy in social networks. In section \ref{sec_evalu}, we evaluate our variation of framework and compare it with one of the state-of-the-art methods. Finally, section \ref{sec_conclud}, concludes this paper.

\section{Related works}
\label{sec_related}

Before being considered from computer science perspective, polarization was a big concern in sociology. Different researches have been conducted to evaluate polarization changes in societies \cite{dimaggio1996have,fiorina2008political,baldassarri2008partisans}. Sociologists evaluate public opinion by quantitative data which is gathered from different surveys. Based on their definition which considers polarization as distance between opponent groups' opinions, sociologists draw their results about the amount of polarity and its change. There are different measures to be used for evaluating quantitative data for calculating polarization. Bramson et al. provided a set of measures to detect opinion distance between two contradicting groups \cite{bramson2016disambiguation}.\\

By the emergence of social networks, computer scientists started to consider this topic from their own perspectives, using new methods of text mining \cite{jain2017extraction}. They tried to overcome the limitations of tradition methods using huge amount of invaluable information provided by users in social networks.\\

Adamic and Glance \cite{adamic2005political} considered blogosphere in US 2004 presidential election. They made their basic network using the established links among the active blogs before the election day and found two well-separated groups of sub-networks related to the Democrats or Conservative. Hargittai et al. \cite{hargittai2008cross} took both quantitatively and qualitatively data into account and found similar behavior in users, like what reported by Adamic and Glance \cite{adamic2005political}. This line was followed by other researchers too; one of the last researches has investigated Brazilian Political Crisis in 2016 \cite{alves2019together}. In the lack of long term research on social polarization in social networks, Garimella et al. \cite{garimella2017long} focused on the long term and persistent topics and provided a long term analysis for polarization in social media. All of the aforementioned studies proved that there exist some kinds of segregations between contradicting political and social groups in social networks, which makes exposure to other side's points of view rare to the users, and this makes the situation even worse, since for alleviating the amount of extremism, contradicting public opinion should be exposed to each other. Table \ref{table1} summarizes the research done  to detect controversy on social media or real life.

\begin{table}
	\caption{Researches for detecting controversy in offline and online communities.}
	\label{table1}
	\centering
	\begin{tabular}{ |c | l | c| c |}
		\hline 
		\multirow{2}{*}{Type} & \multicolumn{3}{c|}{Features} \\
		\hhline{~---}
		&Name & Network & Content\\
		\hhline{----}
		\multirow{4}{*}{Offline}  
		&Dimaggio et al. \cite{dimaggio1996have} & 		- & 	$\times$ \\
		&Fiorina et al. \cite{fiorina2008political} & 		- & 	$\times$ \\
		&Baldassarri et al. \cite{baldassarri2008partisans} & 		- & 	$\times$  \\
		&Bramson \cite{bramson2016disambiguation} & 		- & 	$\times$  \\
		\hline 
		\multirow{4}{*}{Online}   
		&	Adamic et al. \cite{adamic2005political}&		$\times$&		- \\
		&	Hargittai et al.	\cite{hargittai2008cross}&		$\times$&		$\times$\\
		&	Alves	et al. \cite{alves2019together}&		$\times$&		-\\
		&	Garimella	et al.	\cite{garimella2017long}&		$\times$&		-\\
		\hline 
	\end{tabular}
\end{table}
Studies on measuring controversy can be divided into three main groups: Content based, notwork based, and combined models. If an approach uses network science tools, like community detection, random walk, and so on, it is called network based method. If it uses textual or qualitative information analyzing tools, for instance, NLP or statistical metrics, the approach is called content based method. Finally, if an approach uses algorithms for analyzing textual information; at the same time, it uses network science approaches for analyzing network data, the approach is known as a combined method.
\begin{itemize}
	\item Content based: This group of methods go over the content produced by the network's users and measure the controversy using hand labeling, crowd source, and NLP methods (Sociological methods also lay in this group) \cite{dimaggio1996have,fiorina2008political,baldassarri2008partisans,matakos2017measuring,chen2018quantifying,wojatzki2018quantifying,bramson2016disambiguation,al2018studying}.
	\item Network based: This approach utilizes relations made by users and draws the results using only network topological information \cite{garimella2018quantifying,morales2015measuring,guerra2013measure,coletto2017automatic}.
	\item Combined methods: These methods have combined two aforementioned approaches for getting better results \cite{friedkin1990social,degroot1974reaching,matakos2017measuring,chen2018quantifying}.
\end{itemize}

In the rest of this section, we will review proposed algorithms for measuring polarization and controversy in social media.

\subsection{Content Based}
Content based approach includes two sets of methods, which consider contents produced by users in social networks, opinion influence, and survey results. The first class of methods contains sociological perspectives in which after gathering qualitative information from surveys, some statistical methods are used for extracting controversy and bipolarity level in the society. This approach is beyond the scope of our work, so we are not going to investigate it deeply (the interested readers are referred to \cite{dimaggio1996have,fiorina2008political,baldassarri2008partisans,bramson2016disambiguation}). Among these methods, the most similar measure, which has a viewpoint near computer science perspective was proposed by Wojatzk et al. \cite{wojatzki2018quantifying}. They put forward a method for making crowdsourcing claims and ordering them. The claims are contents that collected from the social networks and ranked using the scores assigned by participants. The results are evaluated by NLP methods. The most rudimentary work in this regard is extracting and assigning scores for controversial words and phrases \cite{klenner2014verb,mejova2014controversy}\\

 In the second class, contents are analyzed using the methods provided by NLP methods; these methods similar to the algorithms of the first class are network-free methods, i.e., they do not need network's structural information. Although Garimella et al. \cite{garimella2018quantifying} showed that bag-of-word and sentiment analysis are not distinctive measures for finding controversial issues, Al-Ayyoub et al. \cite{al2018studying} defined some sentiment based methods and proved their usefulness in finding and quantifying controversial topics.\\

Controversy detection is a flourishing topic in the web sphere as well. The main purpose of this branch of controversy detection is making users aware of controversial issues in their web exploration. The articles of wikipedia, the well-known online encyclopedia, have been considered for extracting controversy level of them \cite{kittur2007he,borra2015societal}. The amount of the controversy in the web, also, has been considered deeply \cite{dori2015automated,jang2016improving}. Dori-Hacohen et al. have used the most frequent words of a web page as inquiry for retrieving the wikipedia pages. Next, the controversy scores are assigned to retrieved wikipedia pages, and controversy level of the web page is elicited. Standing on this cornerstone research, Jong et al. \cite{jang2016probabilistic} proposed a probabilistic method not only to find the controversial topics, but also to rank them. Firstly, they provided a probabilistic model \cite{kittur2007he}, and in the next step, a language modeling approach is presented for detecting controversy. Jang et al. \cite{jang2016improving} have remarked two main drawbacks \cite{dori2015automated}, namely, the query problem and score shortcut, and suggested a new method to address these shortcomings. \\

Using sentiment analysis and NLP methods have made the bases of Tsytsarau et al. works \cite{tsytsarau2011scalable} in which a framework for finding contradiction in news articles has been proposed, including three phases: Detecting topics, detecting sentiments, and analyzing the sentiments using mean and variance of the sentiment's distribution.\\

A set of content based researches relies on sentimental signs in documents. Choi et al. \cite{choi2010identifying} considered published articles from June 2001 to May 2002 for extracting controversial topics and their sub-topics. The first step was extracting keywords to be used in the queries. The next step was assigning sentiment score for the negative and positive signals in the documents. Finally, the last step was labeling the extracted topics as controversial or noncontroversial. They also devised a regression model for extracting the subtopics. In similar line, Ignatow et al. \cite{ignatow2016sentiment} used LSA \cite{dumais2004latent} for finding topics of news articles and analyzing them based on sentiment analysis models. \\

Beelen et al. \cite{beelen2017detecting} used articles' comments for finding their controversy. They extracted four sets of features, namely, structural, linguistic, emotional, and WikiPedia similarity. Then, Random Forest and Support Vector Machine were used for predicting controversy news.\\

With regard to the lack of researches considering news articles in addition of social network content, Sriteja et al. \cite{sriteja2017controversy} tried to combine the available data in news media's articles on a topic with social media users reaction toward the topic. Features, such as user opinion, controversial words, and topic intensity, which contain sentiment analysis, lexical analysis, and users interaction, like the number of comment, are taken into account for calculating controversy score.\\

\subsection{Network Based}

In the network based methods, the only instrument to be used is network. These methods purposefully neglect content data as it is deeply context-dependent. Different cultures need various questions in a survey and different NLP methods, and trainers are needed for different languages. In addition, quantifying internal opinions is really a drudgery, if  possible. Instead, these methods rely on structural features of networks, which are context-free and easy to use.\\

Guerra et al. \cite{guerra2013measure} were the first who used topological attributes for extracting controversy level. After finding two disjointed community (for further study of community detection algorithms refer to \cite{papadopoulos2012community}), they divided the nodes into two disjointed groups, the "boundary nodes" and the "internal nodes", the former set includes the nodes, which have at least one edge to other community members (cross-group edges), and the latter set contains the nodes, which do have just edge to the other nodes within their own community (internal edges). The edges that are neither cross-group nor internal are neutral, and then discarded. By defining $d_i$ as the internal degree of a node or number of internal edges and $d_b$ as boundary degree or number of cross-group edges the polarity measures is defined as:
\begin{equation*}
P = \frac{1}{|B|} \sum_{v \in B} \big[ \frac{d_i(v)}{d_b(v) + d_i(v)} - 0.5 \big],
\end{equation*}
where $B$ is the set of boundary nodes. Finally, the authors showed that the polarized topic has low density of high degree nodes in the boundary set. In other words, density of influential users (users with high degree in overall network) is low in the boundary set. Although this method overcomes the drawbacks of modularity measure \cite{newman2006modularity} used in \cite{conover2011political} as polarization measure, it has very limited perspective to the network and considers only a small set of users.\\

Morales et al. \cite{morales2015measuring}, inspired by the electronic dipole moment,  proposed a method similar to the well-known community detection method, label propagation. They found two communities of the network and assigned the most possible amount of controversy (-1 and 1) to the most influential nodes of these communities. Then the process of distributing the assigned amounts of controversy for the selected users continued until convergence. Ultimately, the distance between positive and negative distribution is considered as the controversy measure. The main shortcoming of this method is considering the influencers as the producer of content; however, based on controversy definition from computer science viewpoint, the amount of users' exposure to rival group content should be considered. Consequently, all users should play their own role in the distribution process.\\

The most similar measure to our work is Garimella et al.'s \cite{garimella2018quantifying} method. They also used a community detection method for separating two polar groups (METIS \cite{karypis1997metis}). Then a random walk is used for computing controversy measure. Lets $P_{AB} = P[start in partition A | end in partition B]$ be conditional probability for a random walk; then, RWC (Random Walk) is defined as:
\begin{equation}
	RW = P_{XX}P_{YY} - P_{YX} P_{YX},
\end{equation} 
where $X$ and $Y$ are two communities of network. The random walk starts from each node and continues its journey to reach one of the influential nodes of network from both communities. The authors also presented a measure for each user's controversy. This state-of-the-art method makes clear distinction between controversial and non-controversial issues; however, it suffers from a big shortcoming. The graphs used in this work, similar to the Morales et al. \cite{morales2015measuring}, are simple graph without any extra information on the edges or nodes. Undoubtedly, the presented information in the social networks, regardless of our ability to capture and measure them, can improve the final outcomes. For instance, Castillo et al \cite{castillo2011information} showed that combining social network's information with structural information can improve our ability in detecting piece of fake information in social networks. To the best of our knowledge, the only proposed controversy measuring method for a network with other type of information except the relation is \cite{bonchi2019discovering} which takes into account sign of edges, as well.\\

The entire aforementioned network based methods used a community detection method to segregate two opponent groups. To overcome this problem, Coletto et al. \cite{coletto2017automatic} put forward a motif based and content independent method which, used structural features and motifs for finding controversial issues. This method also built its result only on structural information in the networks.\\

Another usage of network for finding polarization is appeared in Al Amin et al.'s \cite{al2017unveiling} work. The authors used network to reveal polarization too, but their made network is totally from a different nature. A bipartite graph in which one of the sets includes sources or users who have produced a post or retweeted it, and the other set contains the posts that are circulating in the network sphere. Then, using this network a matrix is made in which each entry is the combination of a source activity probability in a camp and a post probability for circulation in it. The next step in this method is factorizing this matrix into two separated matrices, which show the user activity and post circulation probability, separately. This factorization is achieved by optimizing an objective function using gradient decent method. Before this work Akoglu \cite{akoglu2014quantifying} also exploited bipartite graph; however, the used method for extracting controversy is different. Trying to find leaning the users in addition to a measure for ranking their partisan level, a bipartite graph was formed. The members were users and subjects. Markov Random Field was utilized for providing an objective function, and an approximating method for solving the objective function, that is Loopy Belief Propagation, is used.\\

Another opinion based method, which also utilizes the structure of network, is proposed by Amelkin et al. \cite{amelkin2017distance}. Although the method puts forward a measure for calculating individual's polarity level, the main aim of the paper was proposing a distance method for opinion changes in a network. They have practiced transportation problem solution for finding the difference between two sides distribution on the consequent time instance. Their methods are useful for finding anomalies in a network, probably injected by a number of malicious users.\\

\subsection{Combined Models}
The third class has something in common with both network and content based methods. On one hand, they work with content and concepts; on the other hand, they use network for measuring controversy.\\ 
Two of the opinion formation methods \cite{friedkin1990social,degroot1974reaching} have made the foundation of opinion based models. Friedkin \cite{friedkin1990social} proposed a method for opinion formation with two variable, $s_i$ and $z_i$. The $s_i$ is the internal opinion of user $i$ and $z_i$ is the expressed opinion of $i$. Unlike Friedkin's model \cite{friedkin1990social}, in DeGroot's model \cite{degroot1974reaching} user's opinion changes gradually ($s_i$ and $z_i$ play the same role in this method too). Friedkin's model was used further in the controversy detection methods. Matakos et al. \cite{matakos2017measuring} made a network of influence, and then executed a random walk to calculate $z_i$ for each users as "polarization index". Polarization index is used further for evaluating controversy of a network. Friedkin's idea was also used by Chen et al. \cite{chen2018quantifying}. They produced a controversy measure using $\textbf{z}$ internal production ($\textbf{z}$ is vector of all $z_i$s in the network). Our algorithm is different from these opinion based algorithm in the information it utilizes. Our algorithm combines structural information provided by social network (and not opinion network) with user-provided information(and not their opinion).\\

In one of the most recent studies aiming at quantifying controversy in social networks, Bonchi et al. \cite{bonchi2019discovering} made a eigenvector based methods for evaluating controversy in signed graphs. This algorithm is different from the proposed algorithm in this paper in the type of information that it uses. This algorithm only considers sign of edge; however, our algorithm takes all the provided information by the nodes into account.\\

To the best of our knowledge, there is not any algorithm presented to calculate controversy which directly computes controversy level in attributed graphs. Indeed, our approach tries to combine content and structural information, at the same time, provide a flexible framework to be used in different situations and conditions.
\begin{table}
	\caption{Classification of researches for presenting a controversy measure.}
	\label{table2}
	\centering
	\begin{tabular}{ |c | l | c| c |}
		\hline 
		\multirow{2}{*}{Type} & \multicolumn{3}{c|}{Features} \\
		\hhline{~---}
		&Name & Network & Content\\
		\hhline{----}
		\multirow{7}{*}{Sociological}  
		&Bramson et al. \cite{bramson2016disambiguation} & 		- & 	$\times$  \\
		&Dimaggio et al. \cite{dimaggio1996have} & 		- & 	$\times$  \\
		&Fiorina et al. \cite{fiorina2008political} & 		- & 	$\times$  \\
		&Baldassarri et al. \cite{baldassarri2008partisans} & 		- & 	$\times$  \\
		&Wojatzki et al. \cite{wojatzki2018quantifying} & 		- & 	$\times$  \\
		&Klenner et al. \cite{klenner2014verb} & 		- & 	$\times$  \\
		&Mejova et al. \cite{mejova2014controversy} & 		- & 	$\times$  \\
		\hline 
		\multirow{10}{*}{NLP}  
		&Al-Ayyoub et al. \cite{al2018studying} & 		- & 	$\times$  \\
		&Kittur et al. \cite{kittur2007he} & 		- & 	$\times$  \\
		&Borra et al. \cite{borra2015societal} & 		- & 	$\times$  \\
		&Dori et al. \cite{dori2015automated} & 		- & 	$\times$  \\
		&Jang et al. \cite{jang2016improving} & 		- & 	$\times$  \\
		&Jang et al. \cite{choi2010identifying} & 		- & 	$\times$  \\
		&Tsytsarau et al. \cite{tsytsarau2011scalable} & 		- & 	$\times$  \\
		&Ignatow et al. \cite{ignatow2016sentiment} & 		- & 	$\times$  \\
		&Beelen et al. \cite{beelen2017detecting} & 		- & 	$\times$  \\
		& Sriteja et al. \cite{sriteja2017controversy}& 		- & 	$\times$  \\
		\hline 
		\multirow{4}{*}{Network}   
		&	Guerra et al. \cite{guerra2013measure}&		$\times$&		- \\
		&	Morales et al.	\cite{morales2015measuring}&		$\times$&		-\\
		&	Garimella	et al. \cite{garimella2018quantifying}&		$\times$&		-\\
		&	Coletto	et al.	\cite{coletto2017automatic}&		$\times$&		-\\
		&	Al Amin et al. \cite{al2017unveiling}&		$\times$&		-\\
		& Amelkin et al. \cite{amelkin2017distance} &		$\times$&		-\\
		& Akoglu \cite{akoglu2014quantifying} &		$\times$&		-\\
		\hline 
		\hline 
		\multirow{4}{*}{Combined}   
		&	Friedkin \cite{friedkin1990social}&		$\times$&		$\times$ \\
		&	Chen et al. \cite{chen2018quantifying}&		$\times$&	$\times$\\
		&	Bonchi et al. \cite{bonchi2019discovering} &		$\times$&	$\times$\\
		&	BRW &		$\times$&	$\times$\\
		\hline 
	\end{tabular}
\end{table}
\section{Methodology}
\label{sec_method}
In this section, we intend to introduce a method for combining two sources of invaluable information, namely, content or textual data and structural data. We combine these sources of information for broadening horizons of controversy measuring algorithms, and also for surmounting present shortcomings in the proposed algorithms.\\

If $N$ is a network; $N_1$ and $N_2$, where $N_1 \cap N_2 = \emptyset$ and $N_1 \cup N_2 = N$ are two communities of $N$. We want to evaluate the average level of nodes in $N_2$ to which the idea of $i$ as a member of $N_1$ will be exposed.\\

Before introducing our methodology, we provide some definitions, which are going to be used in the rest of this study. \\

\noindent \textbf{Controversial topics:} Controversial topics are the topics, which divide societies into at least two contradicting groups about the issue.\\
\textbf{Attributed graph:} Attributed graph is a graph, which has some extra information for the nodes and/or edges. For instance, number of friends of the user, number of posts, etc. In attributed graph, each node or edge has a vector which contains the corresponding node's information.\\

\subsection{Framework for quantifying controversy in attributed networks}
\begin{figure*}
	\centering
	\begin{subfigure}[t]{0.4\textwidth}
		\includegraphics[width=\textwidth]{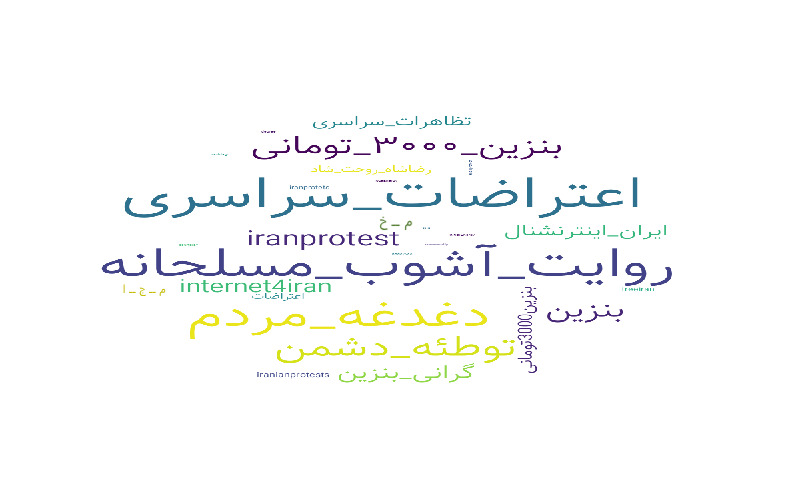}
		\label{wordcloud1}
	\end{subfigure}
	\begin{subfigure}[t]{0.4\textwidth}
		\includegraphics[width=\textwidth]{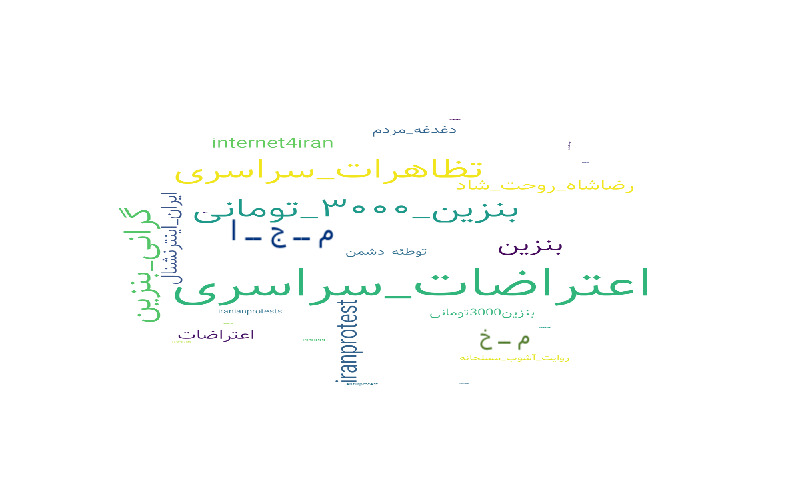}
		\label{wordcloud2}
	\end{subfigure}
	\caption{Word clouds of used hashtags by two contradicting sides of political debates in Persian tweets, November 2019.}
	\label{wordcloud}
\end{figure*}

People's political stance, most of the time, arise from their personal characteristics and psychology. Thus, personal information, which is extractable from users profile, is a dependable source of information. For instance, it is not surprising to see a socialist political activist to support a candidate from the left side of a political spectrum. In addition to the previous information, as user attribute, the used terminology in the produced text and even used hashtags by two sides of a controversial topics shed light on the level of controversy in social network. For instance, Fig. \ref{wordcloud} illustrates the frequency of used hashtags by two sides of a controversial topic (Iran November 2019 protests) in the word clouds. As it is clear, there are some shared hashtags used frequently by two sides, and in the same breath, there are other representative hashtags, which make two sides of a controversial topic distinguishable. For instance, \#3000\_Toman\_pertol is an English equivalent for the frequently used common Persian hashtag among the users in both of the contradicting communities. However, \#riot\_narration or \#people\_mental\_disterbance, as English equivalent to Persian hashtags, are frequently utilized hashtags by pro-government Twitter users. In contrast, protesters frequently used \#m\_j\_E and \#m\_kh, as English equivalent to Persian hashtags, which are the abbreviations of their abusive hashtags toward leaders and governing system. This is the main motivation, which leads us to use user attributes in detecting (and maybe predicting) controversial topics.

Our framework is random walk based approach; however, the random walker does it in a special way. First of all, instead of having an infinitive amount of energy, it has only a special amount of initial energy depending on its starting point's position in the network. Intuitively, this initial energy tries to simulate the possible exposure to an opinion, introduced by a user, in the social network; the more central node, the more ability to be heard. For instance, the probability of being a highly retweeted post or idea, which has been shared by a celebrity is much more than an ordinary user. Therefore, it is reasonable to expect that the amount of being exposed by other users for a special idea, likely depends on the position of producer of the idea. The position of the producer of a node can be calculated by the structural or content features (or both of them) of nodes compressed in a vector.  Additionally, a post or idea do not endlessly circulate in a network and have a lifetime. To take into account these characteristics of the post or idea, we simulate the lifetime of an idea by reducing its initial energy in each step based on the energy loss of the node through which the random walker passes. This is called energy loss of node and is proportionate to the position of a node. Due to being interested in the random walk which passes through the boundary edges, we estimate the position of nodes for calculating energy loss with respect to the average position of the nodes in other communities.\\

Moreover, in Garimella et al.'s algorithm \cite{garimella2018quantifying}, they considered only the random walks, which start from influential users and reach to other influential users on the other sides of a debate. Meanwhile, we believe that the amount of controversy measures should be calculated based on their ability for penetrating the other communities. Networks can be structured in different layer, and depth of each layer can be calculated with respect to its distance from the boundary nodes. Thus, the ability of a random walk in penetrating the community, other than the one which it belongs to, is equal to the maximum layer of nodes (belong to other communities) which it passes through. \\

\begin{figure}
	\centering
	\includegraphics[width=0.7\textwidth]{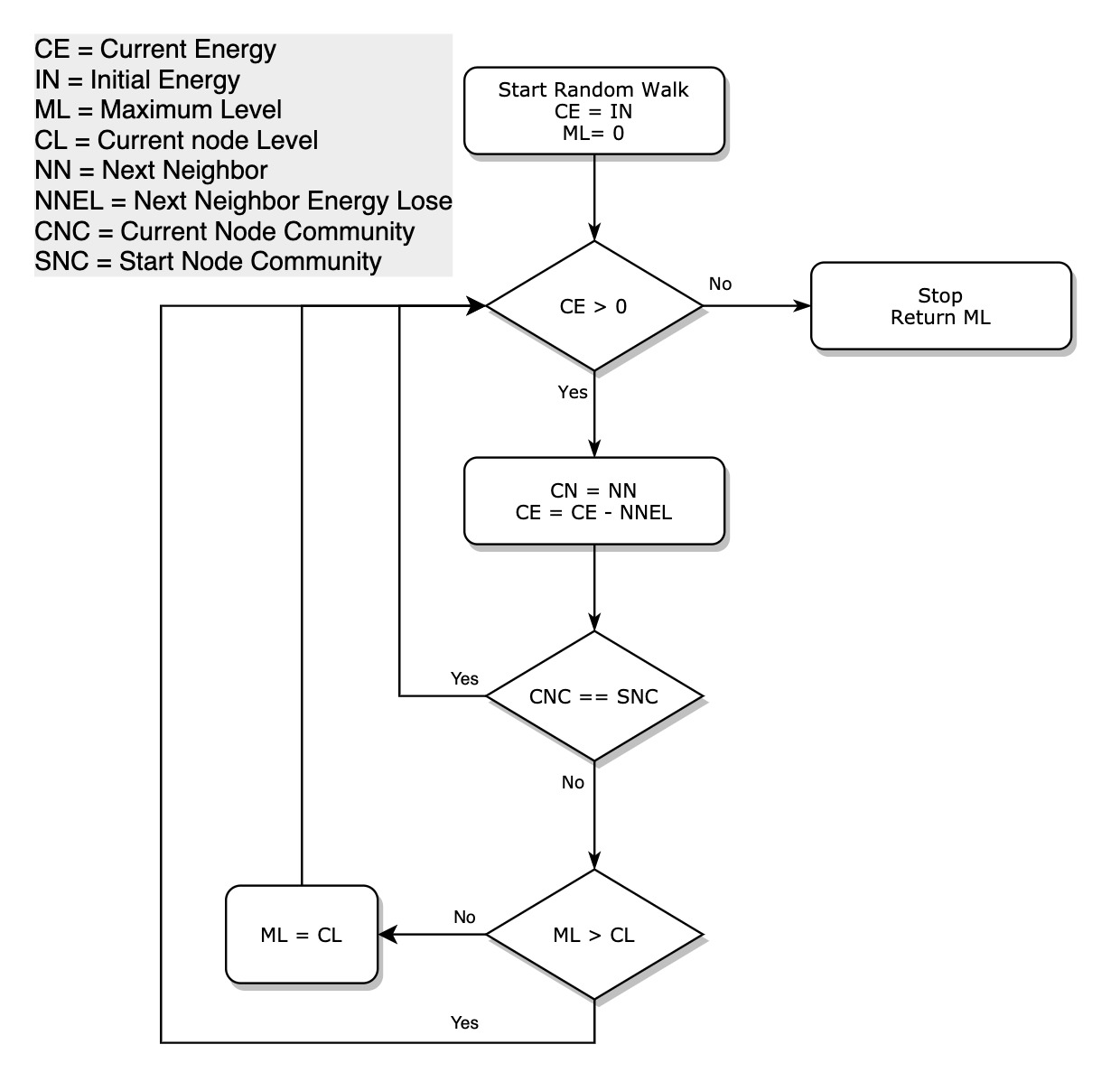}
	\caption{Framework flowchart.}
	\label{flowchart}
\end{figure}

Boundary nodes are defined as the nodes in a community, which have at least one edge to a node in another community. Starting from these boundary nodes and putting their level as zero, the set of level 1 contains the nodes, which only have one or more edges to the nodes in level 0 and nodes in set of level 2. Generally, the level of a node is $n$ if it has at least one link to a node in level $n-1$ and not fewer level, at the same time, at least one link to a node in level $n+1$, if exist (though this definition similar to Breadth First Search (BFS), but it is different in having more than one starting point). Now, we can define a random walker's ability to drilling down the other side's community as the maximum level of nodes, in the other community, through which the random walker passes. Figure \ref{flowchart} indicate the flowchart of this framework.\\

\begin{figure}
	\centering
	\includegraphics[width=0.5\textwidth]{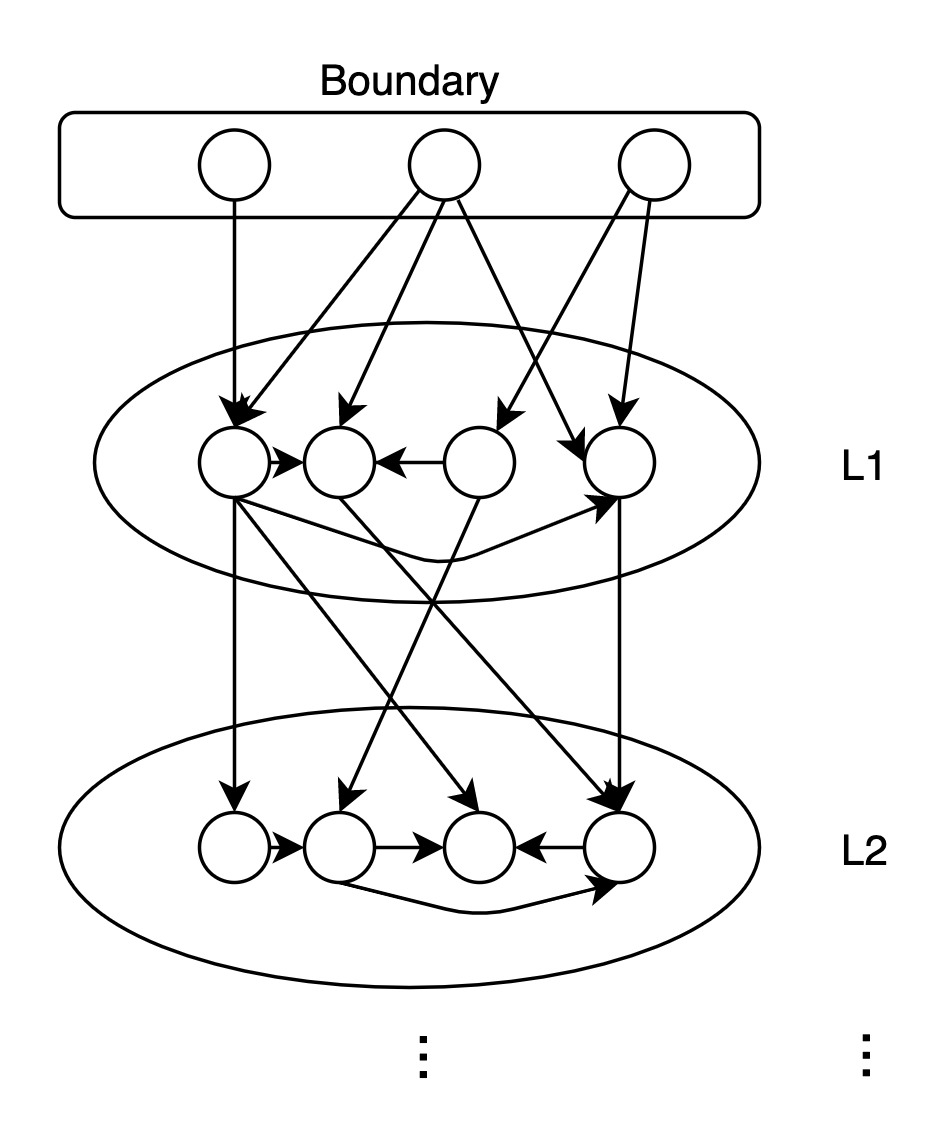}
	\caption{Structure of a community as one side of a debate in a social network.}
	\label{levels}
\end{figure}

To sum up, historically, the controversy level of a topic from a social network perspective has been defined as the average amount of being exposed to contradicting opinion, regardless of corresponding users' reactions (acceptance or rejection) \cite{garimella2018quantifying,guerra2013measure,morales2015measuring}. Users establish closer and stronger echo-chambers for controversial topics. For instance, figure \ref{levels}, indicates a community with a boundary and two levels. In this figure, nodes in level 1 are exposed to information produced by boundary nodes. Thus, the users in level 1 communicate with more like-minded people in comparison to nodes in the boundary. Having propagated a contradicting point of view, boundary nodes have a much moderate viewpoint compared to the nodes in level 1, which only have exposed to the boundary nodes idea. Additionally, the nodes in level 1 have weaker echo-chamber, and therefore, a more moderate point of view compared to the node in level 2, which only sees ideas of the nodes in level 1. Thus, the further from the boundary nodes, the stronger and closer echo-chamber, and the more radical opinion will be.\\

Having this in mind, if an idea is able to go to the deeper levels in the contradicting communities, its controversy level is fewer. Consider \#mothersday and \#gunrights as examples. We know that \#gunrights is much more controversial than \#mothersday. In the case of \#gunrights, and due to its controversy level, the ability of a piece of information to be redistributed by the user with the contradicting point of view is much less than a piece of information containing \#mothersday. This is especially true in deeper levels, which the users in these levels have a more extreme point of view about \#gunrights because of their tighter echo-chamber. Hence, we believe if a random walker, which simulate an idea distribution, reaches to a deeper level in the network, it is less controversial, and vice versa. The average of this ability for all the ideas or tweets regarding a specific topic is overall controversy level of that topic.\\

Hence, the algorithm contains two steps; first, calculating initial energy and energy loss for each node, and next, starting random walks from each node with initial energy, and reducing the initial energy based on the energy loss of the node which the random walker passes through. \\

As the main goal of this study, we aim to determine to what extent an idea produced by a specific user is exposed to the members of contradicting viewpoint. Let's assume that random walker which starts from node $i$ distributes tweet or idea $t$. Then, we have $r_i = (i, t)$, where $r_i$ is the corresponding random walker. From its starting point, $r_i$'s responsibility is simulating the diffusion pattern of $(i,t)$. In each step, $i$ idea's energy is reduced because the probability of being retweeted when a content goes further from its producer decreases. It seems that this energy should be refilled if the content is retweeted by an influencer. However, if $t$ is retweeted by influencer ($j$), then random walk which starts from $j$ and distributes idea $t$, i. e. $r_j = (j,t)$, refills this energy and simulates its distribution pattern. At the same time, $r_i = (i, t)$ as the simulator of $i$'s idea continues its work to death. In fact, the random walk $r_i$ energy should have been retweeted if there were only a set of selected nodes for initiating the random walks, whereas, in our framework, each node has its own random walker. In fact, our algorithm algorithm takes into account influencer and non-influencer's opinion distribution, separately. 

There are several ways to assign initial energy and energy loss in attributed graphs. Here, we introduce two basic methods, which can be freely used in any kind of attributed network; notwithstanding, the framework is flexible enough to be used in different ways.\\

There are several approaches for assigning initial energy for each node, such as degree centrality, betweenness centrality, or eginvector centrality. However, for our version of this framework, we adapt a data mining and machine learning point of view.\\ 

Influencers have higher centrality and more audiences. Therefore, the distance from the central points shows the ability to attract more audiences; these distances determine the initial energy of a node, that is, the energy by which a random walker starts its journey. Furthermore, influencers of the other communities attract more like-mind audiences. Thus, the more distance from other community's influencers, the fewer audiences with the contradicting point of view will be, eventually, the more energy loss for that node we will have. It is worth mentioning that we want to find the ability of a random walker to reach other community; hence, we take the contradicting community's central point into account for calculating energy loss.\\

We have attributes of the nodes; therefore, we can map each of these nodes to a point in the space. We know that these nodes have been assigned to two disjointed sets using a community detection algorithm; hence, we are easily able to find the central nodes of these two node sets using these attributes, which are representative vector of nodes. Having mapped nodes on the space and calculated the central nodes of each group, we define the initial energy and energy loss as follows:

\begin{equation}
\label{initial}
	InitialEnergy = \frac{1}{Dist(node, CC_{in})} + \frac{1}{Dist(node, CC_{out})}
\end{equation}
\begin{equation}
\label{loss}
lossEnergy = Dist(node, CC_{out}) 
\end{equation}

\begin{figure}
	\centering
	\includegraphics[width=0.5\textwidth]{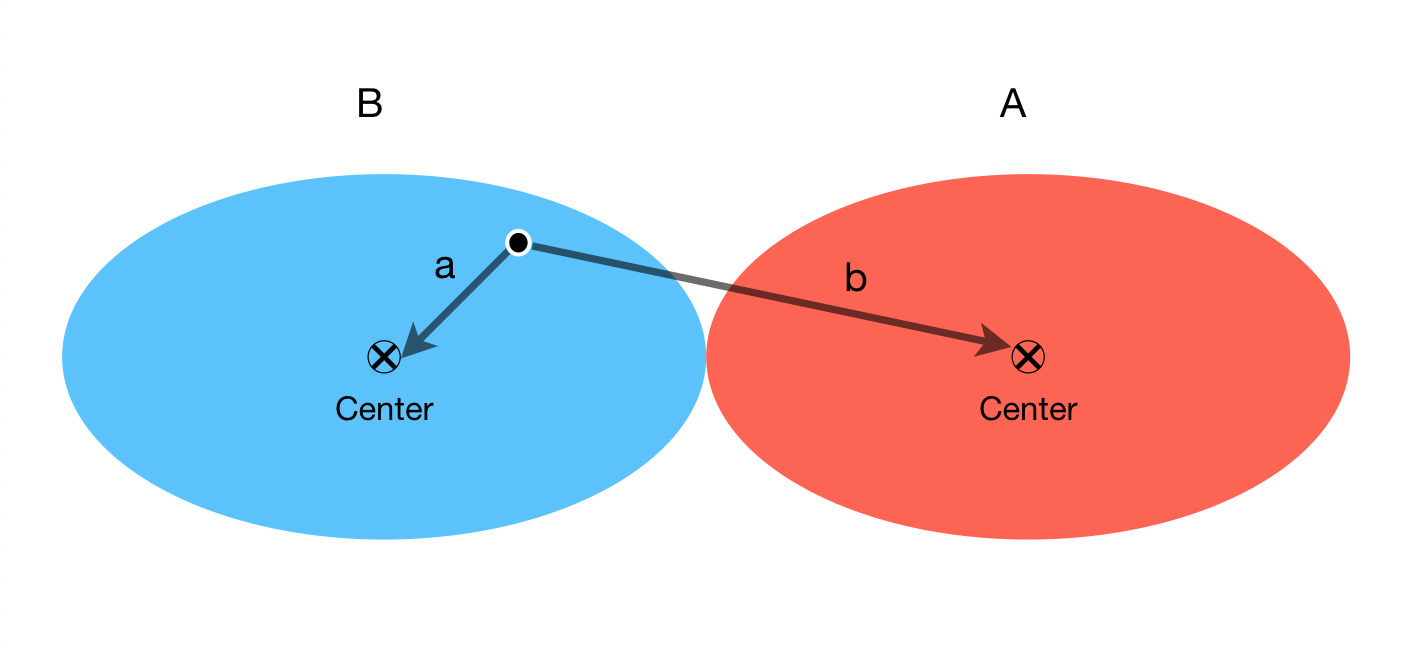}
	\caption{Central and initial points and a node position. Red and blue ellipses are two communities. In this figure we have $InitialEnergy = \frac{1}{a} + \frac{1}{b}$ and $lossEnergy = b $.}
	\label{initial_loss}
\end{figure}

Where $CC_{in}$, $CC_{out}$ are central node of communities, which corresponding node belongs to and does not belong to, respectively, and $Dist(.,.)$ is euclidean distance (or any other distance measure). Intuition for this measure is as follows: the ability to be heard (initial energy) depends on the distance from current community's average point of view to start circulation of a piece of information, and distance from the average point of view (central point of community) of other community (because we are interested in the ability of a random walk in penetration of other community). For energy loss, the intuition is as follows: the farther from middle point of contradicting community, the lower chance to reach to another community, and therefore, to be heard. To avoid trapping in the local neighborhood, the initial energy can be multiplied to a sufficiently big number, but it should be noted that to compare a different network, the same configuration has to be used. Fig. \ref{initial_loss} shows the position and distance of node from central points.\\

In our implementation, especially for being able to compare to other methods, we change network structural information to attributes. In fact, there are several methods for the representation of network \cite{hamilton2017representation}. New representation methods are introduced to map networks to a lower dimension space in which the geometric position of nodes reflect their structural attributes. Therefore, in this representation, we expect to see similar nodes near each other. This similarity in pure networks, without any attributes, is based on the edges; thus, we believe that community nodes should be much closer to each other in newly mapped versions.\\

There are several representation algorithms to map a node on space \cite{grover2016node2vec,tang2015line,perozzi2014deepwalk}. We selected node2vec \cite{grover2016node2vec} to map our network to space. The first objective of node2vec is preserving network's neighborhood structure, and it performs this using a biased random walk. Having this in mind, we try to assign a vector to each node, and this vector is the output of node2vec algorithm. Now, we have structural information of a node as its attribute.\\

Additionally, we have other attributes as vector for nodes. By concatenating on these features, such as used hashtag frequency or TF-IDF of used words, we can have broader perspective toward the node.\\

However, we have more things to do for achieving a better loss function. As mentioned before, we have two sets of nodes: Boundary nodes and non-boundary nodes (which we call them internal nodes). Our loss function, for simulating the ability of a node in transferring a piece of information from belonging community to other community, has two kinds of perspectives, local perspective and global perspective. In local perspective, we just consider position of a node in the corresponding community with its neighbors. Two different measures for two different kinds of nodes, boundary and internal, are available. First, conductance \cite{yang2015defining} for a node is defined as the ratio of of edges, which stay inside the community, to the edges which go to the other community:
\begin{equation}
\label{conduct}
	Cond_i = \frac{|edge_{out}|}{|edge_{in}|},
\end{equation}
where  $|edge_{out}|$, $|edge{in}|$ are the number of edges, which go out of the community that node $i$ belongs to and the number of edges that stay inside the community, which node $i$ belongs to, respectively. Concentrating on the definition of internal nodes which do not have any links to a node belonging to another community, it is evident that conductance is only meaningful for the boundary nodes and is equal to zero for internal nodes. Indeed, conductance is the ability of a node for passing a piece of information to another community. For internal nodes, this ability lays on the level to which the node belongs. For example, ability of a node which is in the level $n$ for passing a piece of information to other community is more than a node which lay in level $n+1$ since it is a level far from a node in another community. Thus, we need another measure for internal nodes, and it is exactly the level to which a node belongs. We call this measure closeness to demonstrate how much a node is close to another node with different community:
\begin{equation}
\label{closeness}
colseness_i = level_i,
\end{equation}
where $level_i$ is the level which node $i$ belongs to.\\

For global perspective, we regard the distance of a node from the central node of another community using attributes of the nodes represented by vectors. The more distance to contradicting community's central point, the more loss in energy, and the less ability to pass a piece of information to other side:
\begin{equation}
\label{node_distance}
dist2center_i = Dist(i, CC_{out}).
\end{equation}
We know that distance from the other community's central point, and closeness has direct relation with energy loss, and conductance have inverse relation with it. Finally, we have this relation for the loss function:
\begin{equation}
\label{loss_function}
loss_i = \alpha*\frac{closeness_i}{max\_close} + \beta*dist2center_i
 + (\gamma * \frac{max\_cond - Cond_i}{max\_cond})
\end{equation}
where $max_cond$ is maximum conductance in the graph and $\frac{max\_cond - Cond_i}{max\_cond}$ is used for normalization. $max\_close$ is the maximum closeness for the nodes whose corresponding node is in the same community with them. $\alpha, \beta, \gamma$ can be used for tuning the effect of each elements (we put all of them equal to 1 in our version).\\

Algorithm \ref{algorithm_framework} shows how the Biased Random Walk works. In each step, if the current energy is more than 0 one of the neighbors is chosen as target node and the current energy will be reduced proportionate to energy loss of the target node. It is worth to mention that due to being direct graph, it is not surprising to see some nodes in each level as dangle nodes, which do not have any outgoing edge. In these cases for the next step, the BRW choose one of the node in the same level as the current node, then, the current energy is reduced proportionate to target node energy loss, and BRW continue its work to run out its energy ($current_energy < 0$).
\begin{algorithm}[]

	\KwData{An attributed graph}
	\KwResult{Maximum level of a node, belonging to other side of debate, which random walk pass through.}
	Pre-processing: Assign Initial energy and energy loss for each node\;
	MaxList = []\;
	\While{There is an unprocessed node}{
		Maximum level = 0;
		Current Node = Starting Node of a random walk;
		Current energy = Initial Energy of Current Node;
		\While{Current energy> 0}{
			current node = randomly selected neighbor of current node\;
			Current energy = Current energy -  current node's energy loss;
			\If{Maximum level < Current node's level}{
				Maximum level = Current node's level / maximum level of community;
			}	
		}
		Add Maximum level to MaxList
	}
	return MaxList\;

	\caption{Framework for calculating controversy level of an attributed graph.}
	\label{algorithm_framework}
\end{algorithm}

The next step is extracting the ability of each random walker in penetration the other community. After starting the journey from a node, the maximum level of nodes, laying in other community, divided into the maximum level of that community can be considered as the random walker's exposure level. There are two versions of this exposure level.
\begin{itemize}
\item First, the average on all  journeys, regardless of their ability in crossing boundaries or staying on current community, (in this version if random walk stays in its own community the exposure level is equal to zero. If it crosses the boundary nodes the exposure level is one unit more than the maximum level of node through which the random walker passes, for making this controversy level distinguishable from the previous condition by assigning 1 to the boundary nodes of other community). We call this measure which is bigger for non-controversial topics and smaller for controversial topics as \textbf{R}andom \textbf{W}alk \textbf{P}enetration \textbf{R}ate (RWPR).

\item Second, the average only on the random walks which crosses the boundary of communities (in this version, as well, the exposure level is one unit more than node's level), known as \textbf{B}oundary \textbf{C}rossed \textbf{R}andom \textbf{w}alk \textbf{P}enetration  \textbf{R}ate (BCRPR). For considering both of these measures, we calculate RWPR and BCRPR \textit{average}, too.
\end{itemize} 

It is clear that our implementation of this framework is not the only one and each user based on her or his needs can adapt the framework and have its own version of this framework. In the evaluation section, we also concatenate on textual attributes to the structural attributes vector to evaluate the other attribute role in finding controversy.
\textbf{Complexity.} For each node a predefined number of walks, $w$, start. Each of the walks starts with a specific start energy, which depends on the start node's features and other nodes' position in the space; we call average start energy of all nodes as $\bar{s}$. Then, the walk goes forward until it run out of energy, and this depends on loss energy of nodes. We call average loss energy of network as $\bar{l}$. Now, average complexity of algorithm is:
\begin{equation}
	O(BRW) = nw\frac{\bar{s}}{\bar{l}},
\end{equation}
where $n$ is number of nodes in the network.
\section{Evaluation}
\label{sec_evalu}
In this section, we are going to evaluate our implementation of framework (BRW). Two kinds of evaluations are used to show the capabilities of BRW. First, to show why we need new information source to get more reliable results, we compare BRW with RW \cite{garimella2018quantifying}, which has been shown to outperform all other state-of-the-art algorithms because we perform our evaluation on the data used in \cite{garimella2018quantifying} and in that work, the authors showed that their measure outperform other methods, we only compare our methods with RW. Although our methods, unlike other modern methods, are not directly presented for structural features, the results are not worse than new random walk based model \cite{garimella2018quantifying}. In addition, we deeply analyze the behavior of the BRW using the collected data from Persian tweets to show how changes in structural and content information are reflected in the outcomes of BRW.

\subsection{Biased random walk Vs. Pure random walk}
\label{subsub}
As mentioned in section \ref{sec_related}, one of the last methods for evaluating controversy level of a topic is RW based method, which Garimella et al. introduced in \cite{garimella2018quantifying}. This method has been presented to directly deal with structural features of a network. In this section, we compare our methods with the RW. Even though we overlook some precious information by compressing the structural information in a much smaller vector using Node2vec, our results in detecting controversial topic is not worse than RW. Indeed, the RW, like our method, is not able to be more efficient than random choice in detecting controversial topics, based on the data collected by Garimella et al. \cite{garimella2018quantifying}, this is one of our motivations in considering another source of information.\\

\subsubsection{Data set}
For assessing our method with Random Walk, we use the dataset presented in \cite{garimella2018quantifying}. The datasets contain follower-followee network in addition to retweet network; however, there is a consensus in the literature, which the best network for detecting controversial topics is retweet network \cite{conover2011political}. Therefore, we only work with retweet network which is shown in the Table \ref{tab1} and Table \ref{tab2} as controversial and non-controversial networks, respectively\footnote{We have used these network as it has been presented; however, there is only one exception. Mothers' day network was prohibitively big, to deal with this problem we took half of the nodes randomly and made their induced graph.}.\\

\begin{table*}
	\centering
	\caption{Information of controversial topics networks \cite{garimella2018quantifying}.}
	\label{tab1}
	\begin{tabular}{|l r r r l|}
		\hline
		Hashtag & \#Tweets & \#Node & \#Edge & Description and collection period (2015)\\
		\hline
		\#beefban & 422 908 & 21 590 & 30 180 & Government of India bans beef, Mar 2–5\\
		\hline
		\#nemtsov & 371 732 & 43 114 & 77 330 & Death of Boris Nemtsov, Feb 28–Mar 2\\
		\hline
		\#netanyahuspeech & 1 196 215 & 122 884 & 280 375 & Netanyahu's speech at U.S. Congress, Mar 3–5\\
		\hline
		\#russia\_march & 317 885 & 10 883 & 17 662 & Protests after death of Boris Nemtsov ("march"), Mar 1–2\\
		\hline
		\#indiasdaughter & 776 109 & 68 608 & 144 935 & Controversial Indian documentary, Mar 1–5\\
		\hline
		\#baltimoreriots & 1 989 360 & 289 483 & 432 621 & Riots in Baltimore after police kills a black man, Apr 28–30\\
		\hline
		\#indiana & 972 585 & 43 252 & 74 214 & Riots in Indiana pizzeria refuses to cater gay wedding, Apr 2–5\\
		\hline
		\#ukraine & 514 074 & 50 191 & 91 764 & Riots in Ukraine conflict, Feb 27–Mar 2\\
		\hline
		\#gunsense & 1 022 541 & 30 096 & 58 514 & Gun violence in U.S., Jun 1–30\\
		\hline
		\#leadersdebate & 2 099 478 & 54 102 & 136 290 & Debate during the U.K. national elections, May 3\\
		\hline
	\end{tabular}
	
\end{table*}
\begin{table*}
	\centering
	\caption{Information of non-controversial topics' network \cite{garimella2018quantifying} networks}
	\label{tab2}
	\begin{tabular}{|l r r r l|}
		\hline
		Hashtag & \#Tweets & \#Node & \#Edge & Description and collection period (2015)\\
		\hline
		\#sxsw & 343 652 & 9304 & 11 003 & SXSW conference, Mar 13–22\\
		\hline
		\#1dfamheretostay & 501 960 & 15 29 & 26 819 & Last OneDirection concert, Mar 27–29\\
		\hline
		\#germanwings & 907 510 & 29 763 & 39 075 & Germanwings flight crash, Mar 24–26\\
		\hline
		\#mothersday & 1 798 018 & 155 599 & 176 915 & Mother's day, May 8\\
		\hline
		\#nepal & 1 297 995 & 40 579 & 57 544 & Nepal earthquake, Apr 26–29\\
		\hline
		\#ultralive & 364 236 & 9261 & 15 544 & Ultra Music Festival, Mar 18–20\\
		\hline
		\#FF & 408 326 & 5401 & 7646 & Follow Friday, Jun 19\\
		\hline
		\#jurassicworld & 724 782 & 26 407 & 32 515 & Jurassic World movie, Jun 12-15\\
		\hline
		\#wcw & 156 243 & 10 674 & 11 809 & Women crush Wednesdays, Jun 17\\
		\hline
		\#nationalkissingday & 165 172 & 4638 & 4816 & National kissing day, Jun 19\\
		\hline
	\end{tabular}
	
\end{table*}
In these datasets, the hashtags are the keywords for collecting data from Twitter. Table \ref{tab1} contains topics which are controversial, and Table \ref{tab2} includes the information of non-controversial topics (In fact, we tried to extract textual information of tweets because the tweet IDs were available; however, we could only collect about 30 percent of tweets due to being deleted or private account. Therefore, the information was not enough to be used in the content feature based evaluation). 
\subsubsection{Results}
The final results of controversy level for RW and RWPR for BRW is shown in Table \ref{tab3} and Table \ref{tab4} for controversial and non-controversial topics, respectively. We used published code of RW, by the authors\footnote{https://github.com/gvrkiran/controversy-detection/archive/master.zip}, for this assessment. The default configuration was used for RW in which the random walk repeated for 100 times. There are much more things to do in our algorithm. However, for simplicity, we tried to keep everything as simple as possible. We used node2vec for getting embedded vector of each node. The random walk repeated 50 times for each node, and in this special case, we do not multiply the initial energy to a big number.\\

\begin{figure*}
	\centering
	\begin{subfigure}[t]{0.7\textwidth}
		\includegraphics[width=\textwidth]{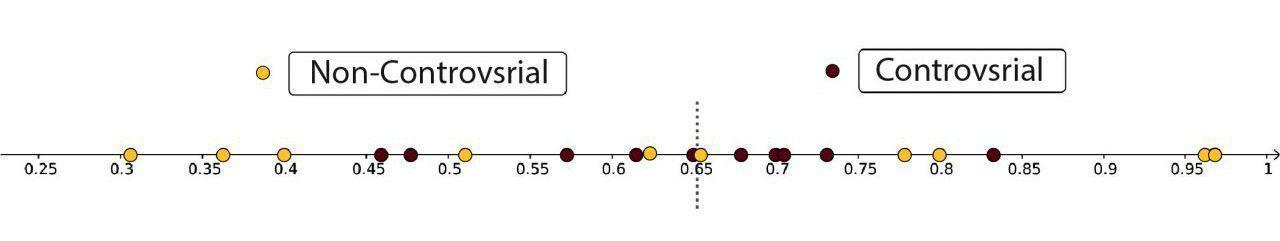}
		\caption{Controversy measure for RW.}
		\label{fig:gull}
	\end{subfigure}
	\begin{subfigure}[t]{0.7\textwidth}
		\includegraphics[width=\textwidth]{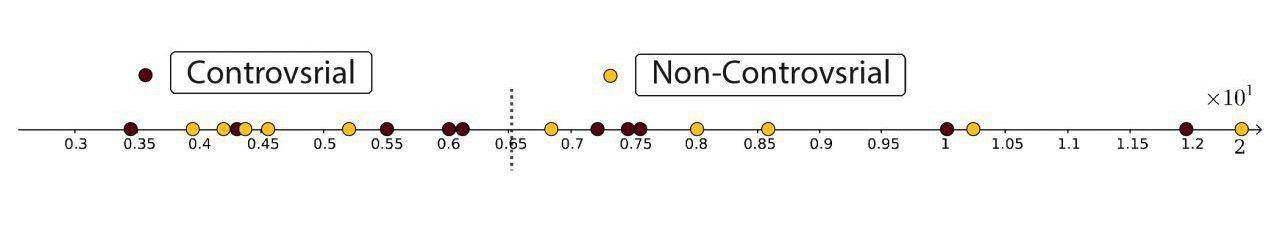}
		\caption{Controversy measure for BRW.}
		\label{fig:tiger}
	\end{subfigure}
	\caption{Controversy measure for RW and BRW.}\label{fig:animals}
	\label{fig_1}
\end{figure*}

In Fig. \ref{fig_1}, we show scatter plots of controversy measure for RW and BRW. There are 10 controversial (red) and 10 non-controversial (blue) topics in each plot. After calculating the measures, we separated 10 first points (10 second points)\footnote{The penetration level for BRW, unlike controversy level for RW, is big when topic is not controversial.} in RW (BRW) plot as non-controversial topics, and the rest as controversial topics. As it is evident both of the methods are not able to detect controversial topics better than chance, precision of both of the methods are 0.5. However, it should be noted that our method is not an only structural based method (unlike RW), and during the compression process (using node2vec), we have to neglect some information. This poor performance pushes us toward taking into account other available sources of information in the networks.

\begin{table}{t}
	\centering
	\caption{RW and BRW for controversial topics.}
	\label{tab3}
	\begin{tabular}[\columnwidth]{|l c c|}
		\hline
		Hashtag &Random Walk & BRW \\
		\hline
		\#beefban & 0.7374 & 0.0731 \\
		\hline
		\#nemtsov & 0.6879 & 0.0631 \\
		\hline
		\#netanyahuspeech & 0.5804 & 0.0601 \\
		\hline
		\#russia\_march & 0.8204 & 0.0755 \\
		\hline
		\#indiasdaughter & 0.6117 & 0.0765 \\
		\hline
		\#baltimoreriots & 0.7014 & 0.1287 \\
		\hline
		\#indiana & 0.4662 & 0.1044 \\
		\hline
		\#ukraine & 0.6522 & 0.0432 \\
		\hline
		\#gunsense & 0.7095 & 0.0556 \\
		\hline
		\#leadersdebate & 0.4632 & 0.0359\\
		\hline
	\end{tabular}
	
\end{table}
\begin{table}
	\centering
	\caption{RW and BRW for non-controversial topics.}
	\label{tab4}
	\begin{tabular}[\columnwidth]{|l c c|}
		\hline
		Hashtag &Random Walk & BRW \\
		\hline
		\#sxsw & 0.8062 & 0.122 \\
		\hline
		\#1dfamheretostay & 0.5157 & 0.0426 \\
		\hline
		\#germanwings & 0.6577 & 0.0441 \\
		\hline
		\#mothersday & 0.4012 & 0.0684 \\
		\hline
		\#nepal & 0.6298 & 0.0457 \\
		\hline
		\#ultralive & 0.3761 & 0.0396 \\
		\hline
		\#FF & 0.9697 & 0.0820 \\
		\hline
		\#jurassicworld & 0.7775 & 0.0539 \\
		\hline
		\#wcw & 0.9627 & 0.0873 \\
		\hline
		\#nationalkissingday & 0.3193 & 0.2064 \\
		\hline
	\end{tabular}
	
\end{table}

\subsection{Iran political environment under consideration}
Iran has been one of the most bipolar societies in the world due to the evergrowing gap between government and a considerable portion of the people, and this makes this country a good source of data for controversial topics. Over the past few years, there have been a number of protests and riots for different reasons. One of the last protests embarked because of a great increase in the petrol price in November 15th of 2019 initiated a chain of protests and riots in  different cities of Iran. After a while, this gap has been filled as an aftermath of Iran popular general's death in January fifth of 2020 by the United State government. We collected Persian tweets from 1 November 2019 to 1 February 2020 for covering all the important events in Iran political environment. We extracted different tweets for different aims, which are elaborated upon in the rest of this section.
\subsubsection{Iran November protests}
In the midnight of November 15th the government suddenly announced that the price of petrol would increase from the next day. Iranians who had been living in a hard life under the United State sanctions started a chain of protests on the day after this announcement. The protests lasted for several days and a number of people were killed in the riots. We collected Persian tweets 15 days before and 15 days after this event. As a preprocessing phase, we extracted all the related hashtags to this event (political hashtag with more than one occurrence), and using these hashtags all tweets containing these hashtags were elicited. After establishing the corresponding network, the largest connected component containing 1216 nodes and 2676  edges was used as the network for assessment.\\

 For extracting the attributes of nodes, we only used existing hashtags in the tweets of each user (other information like profile information and word TF-IDF also can be used, but we restricted our data to hashtags). For each user, there was a table with two rows, the first one containing the list of used hashtags by all users, and the second row, including the number of times the corresponding hashtags had been used by the user.\\

In this section of evaluation, we are going to show how RWPR changes when we impose some noise on the node's attributes and structure. The imposed noises are:
\begin{itemize}
	\item Structural noise: With a probability for each node, the noise adds an edge from the corresponding node to a random node in the graph.
	\item Attribute noise: With a probability in this kind of noise, the hashtag table of corresponding node is replaced with the hashtag table of another randomly selected node.
\end{itemize}

For clearly assessing the behavior of BRW, we made three versions of network for this data set. In the first version, we only took into account the structural features of network, extracted by \textbf{Node2vec}. Indeed, Node2vec was used to produce a vector with 20 dimensions as the features of nodes. In the second version, hashtags were the only feature of nodes. For having equal dimension with Node2vec version, we utilized PCA to reduce the dimension of hashtag table for each user to be 20, known as \textbf{hashtag} network. Finally, we produced a vector with 10 elements for each user, provided by Node2vec, and put it beside the reduced vector (with 10 elements) of hashtags to make a vector with 20 dimensions, we call this version \textbf{both}.\\

As mentioned before, we have three measures: RWPR, BCRPR, and  RWPR and BCRPR average ( for shortening average). We illustrate these measures' changes under different circumstances.\\

\begin{figure*}
	\centering
	\begin{subfigure}[t]{0.3\textwidth}
		\includegraphics[width=\textwidth]{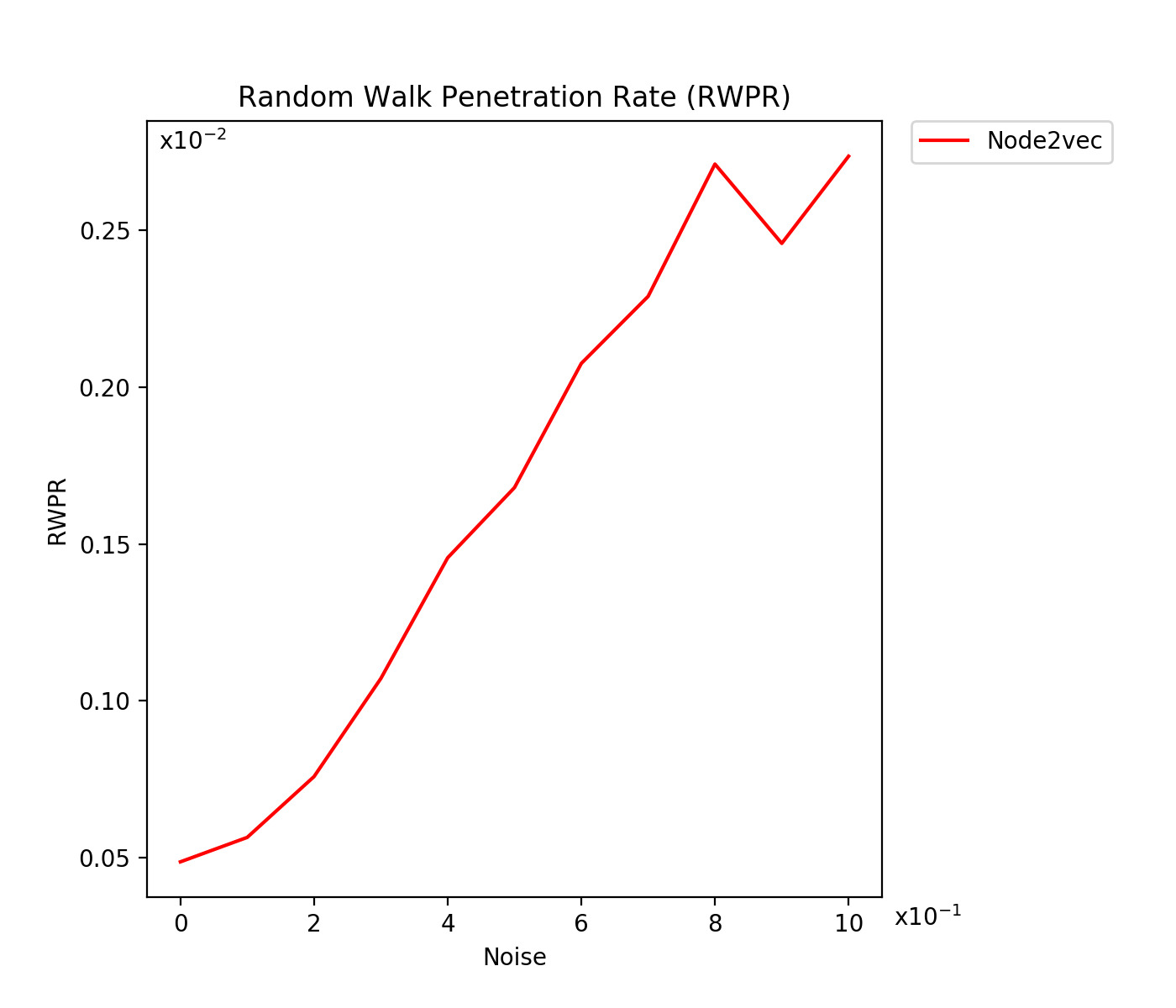}
		\caption{RWPR of Node2vec network. }
		\label{Figure_4}
	\end{subfigure}
	\begin{subfigure}[t]{0.3\textwidth}
		\includegraphics[width=\textwidth]{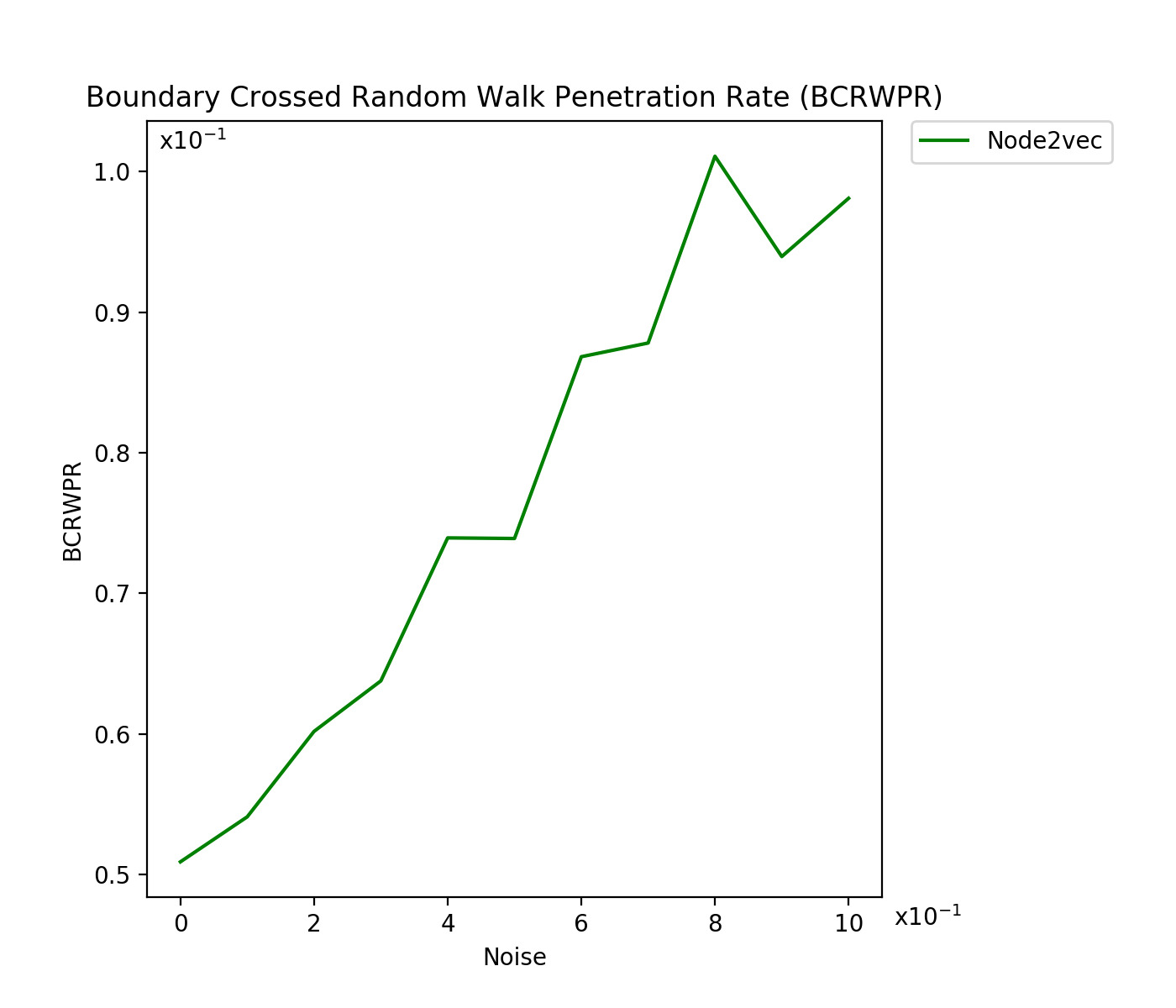}
		\caption{BCRPR of Node2vec network.}
		\label{Figure_5}
	\end{subfigure}
	\begin{subfigure}[t]{0.3\textwidth}
		\includegraphics[width=\textwidth]{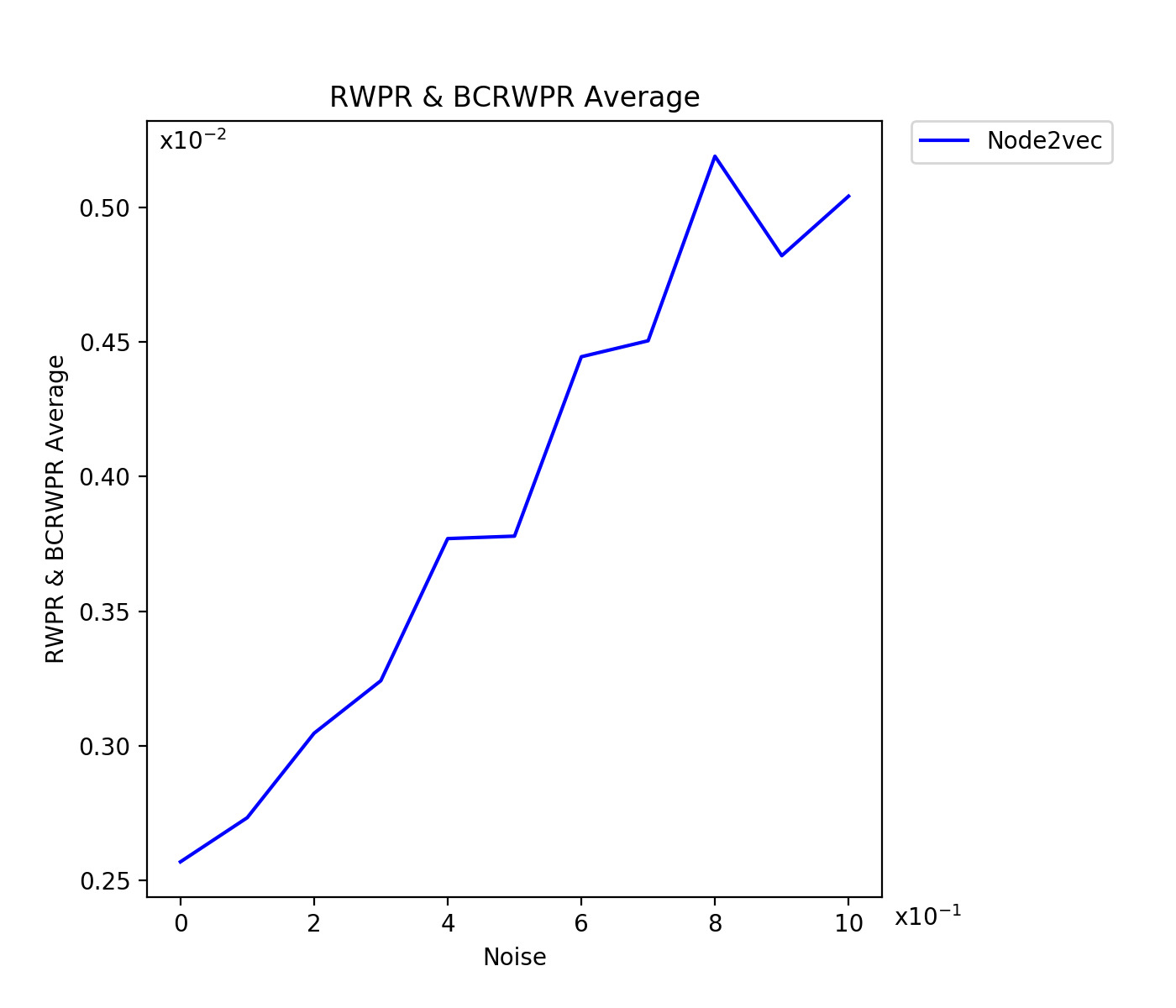}
		\caption{Average of RWPR and BCRPR.}
		\label{Figure_6}
	\end{subfigure}
	\caption{Graph of BRW behavior with different level of noise for Node2vec network. These graphs indicate reduction in the controversy level of network by injecting noise to the network.}
	\label{Figure_node2}
\end{figure*}

First of all, we consider Node2vec version \ref{Figure_node2}. Clearly, the amount of RWPR, BCRPR, and \textit{average} increase by the increase of noise. This demonstrates that the ability of penetration increases as the number of randomly assigned edges rise. We know that one of the main problems of controversial topics is the lack of boundary nodes to help the diffusion of information from one side to another side. The more randomly added edges, the more boundary nodes, and the bigger RWPR will be. As we see below, the only clear increasing behavior for BCRPR is in Node2vec version because randomly added edges by connecting nodes from different level compacts, and the depth of the graph make the nodes in contradicting communities near to each other; therefore, the crossed random walk can easily achieve the farther level of contradicting community.\\

\begin{figure*}
	\centering
	\begin{subfigure}[t]{0.4\textwidth}
		\includegraphics[width=\textwidth]{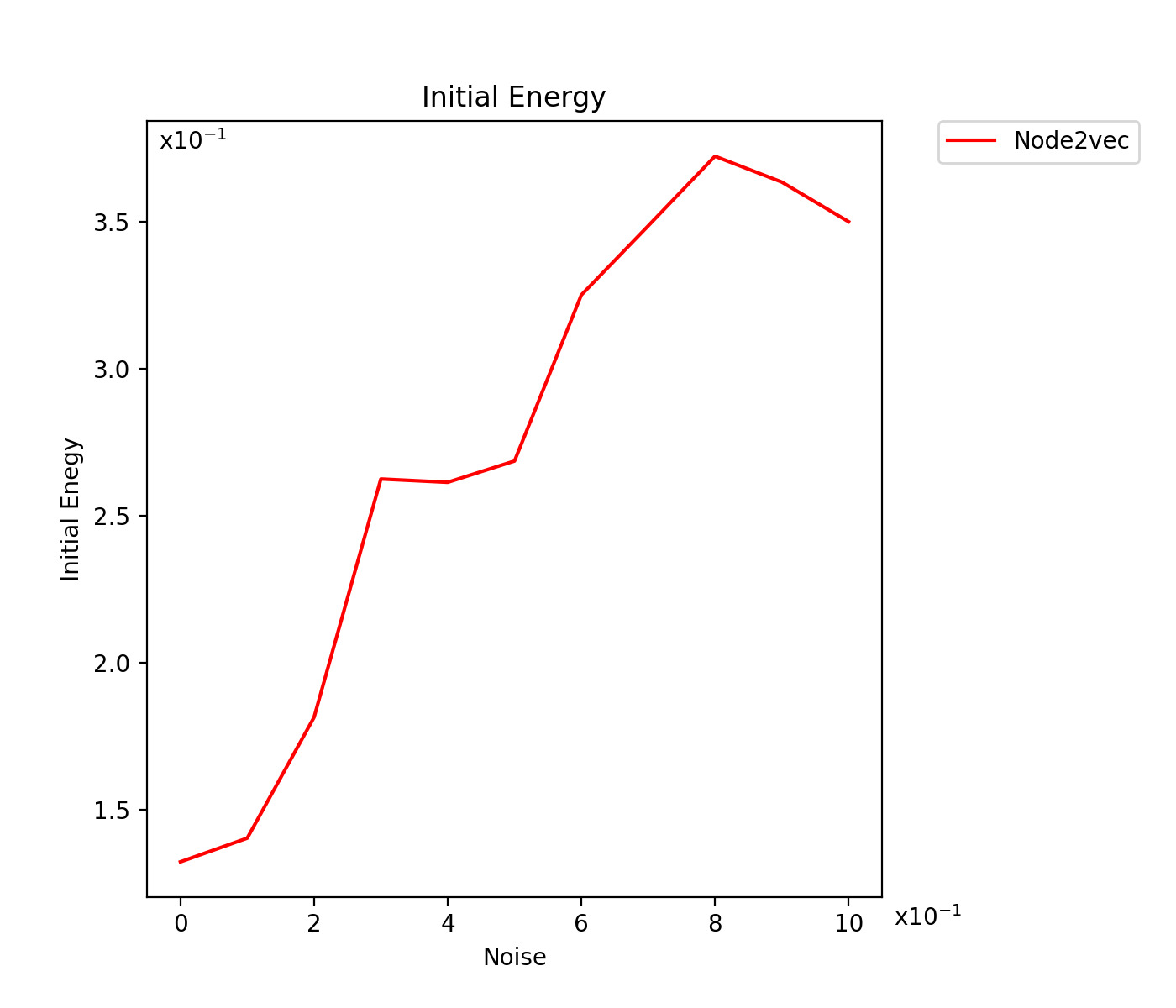}
		\caption{Average initial energy of Node2vec network for different level of noise.}
		\label{Start_2}
	\end{subfigure}
	\begin{subfigure}[t]{0.4\textwidth}
		\includegraphics[width=\textwidth]{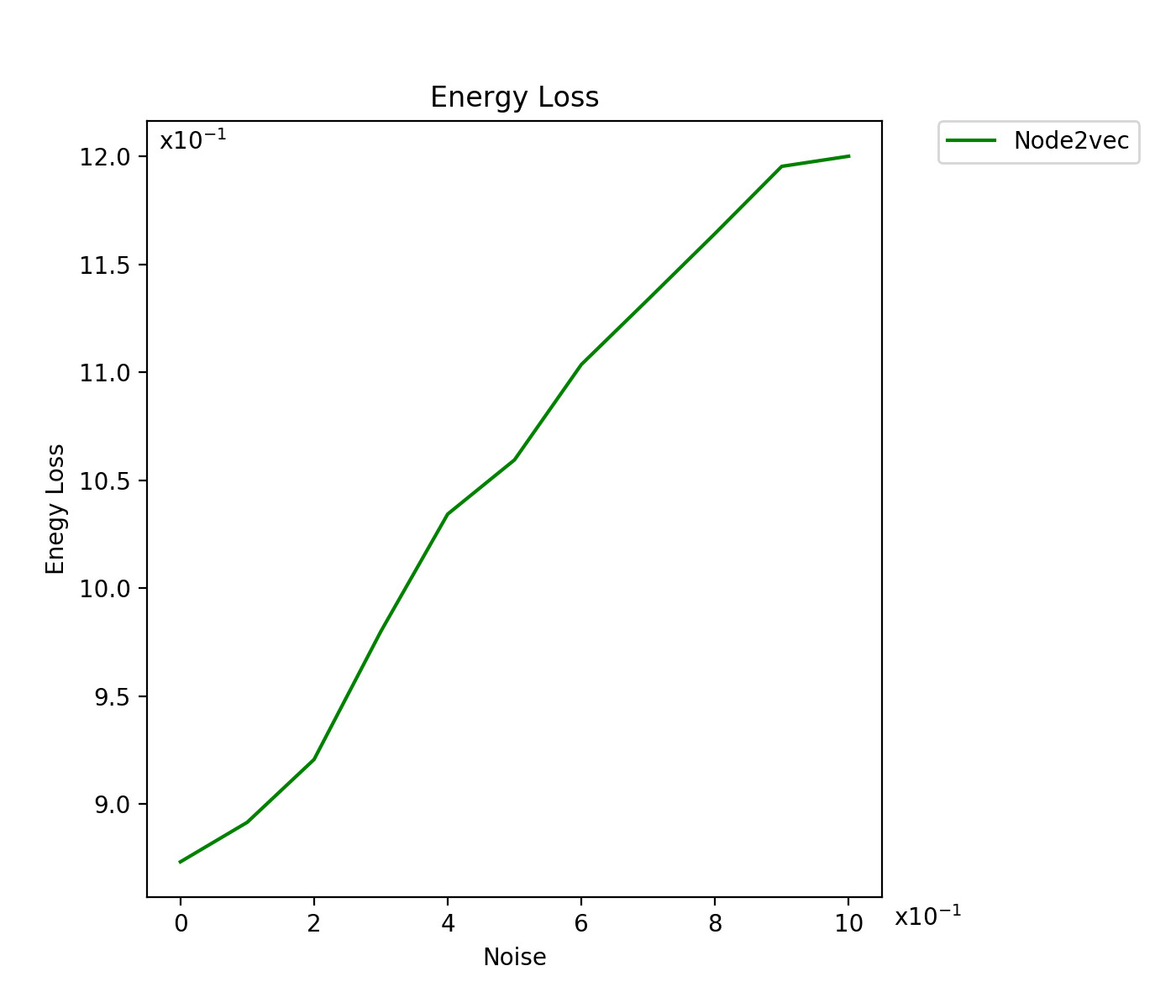}
		\caption{Average energy loss of Node2vec network for different levels of noise.}
		\label{loss_1}
	\end{subfigure}
	\caption{Average initial energy and energy loss in Node2vec network. The graphs demonstrate that central points of communities by injecting noise approach to each other. Hence, we have increase in the average initial energy.}
	\label{intital_node2}
\end{figure*}

\begin{figure}
	\centering
		\includegraphics[width=0.7\columnwidth]{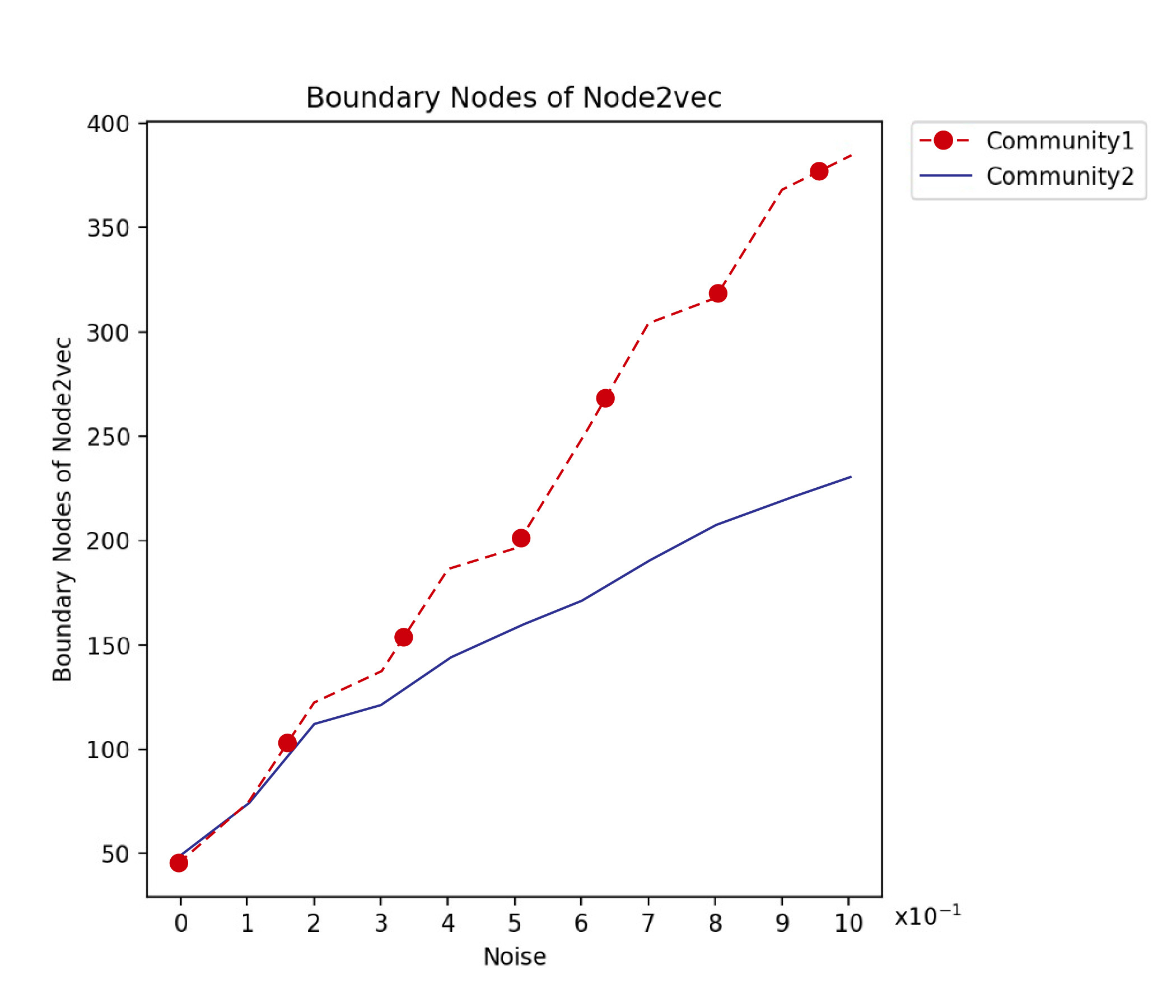}
		\caption{Number of boundary nodes in Node2vec graph for different level of noise. When we inject noise to the network some of the nodes from different communities connect to each other and join to boundary node's set. Therefore, number of boundary nodes increase.}
		\label{Figur}
\end{figure}

Fig. \ref{intital_node2} indicates the average initial energy and energy loss in Node2vec network. Although initial energy's behavior is in line with what we expect, the energy loss increases. It seems that this behavior stems from the dissimilarity imposed by new edges among the nodes. However, this increase in the amount of energy loss can be compensated by the increase in the number of boundary nodes. We know that boundary nodes are the nodes which play as bridges between two communities. New randomly added edges increase the number of crossing edges between two communities; consequently, increase the number of boundary nodes. The more boundary nodes we have, the more random walkers which go through one side to other side of a network will be. Fig. \ref{Figur} shows the increase in number of boundary nodes which compensates the increase in energy loss.

\begin{figure*}
	\centering
	\begin{subfigure}[t]{0.3\textwidth}
		\includegraphics[width=\textwidth]{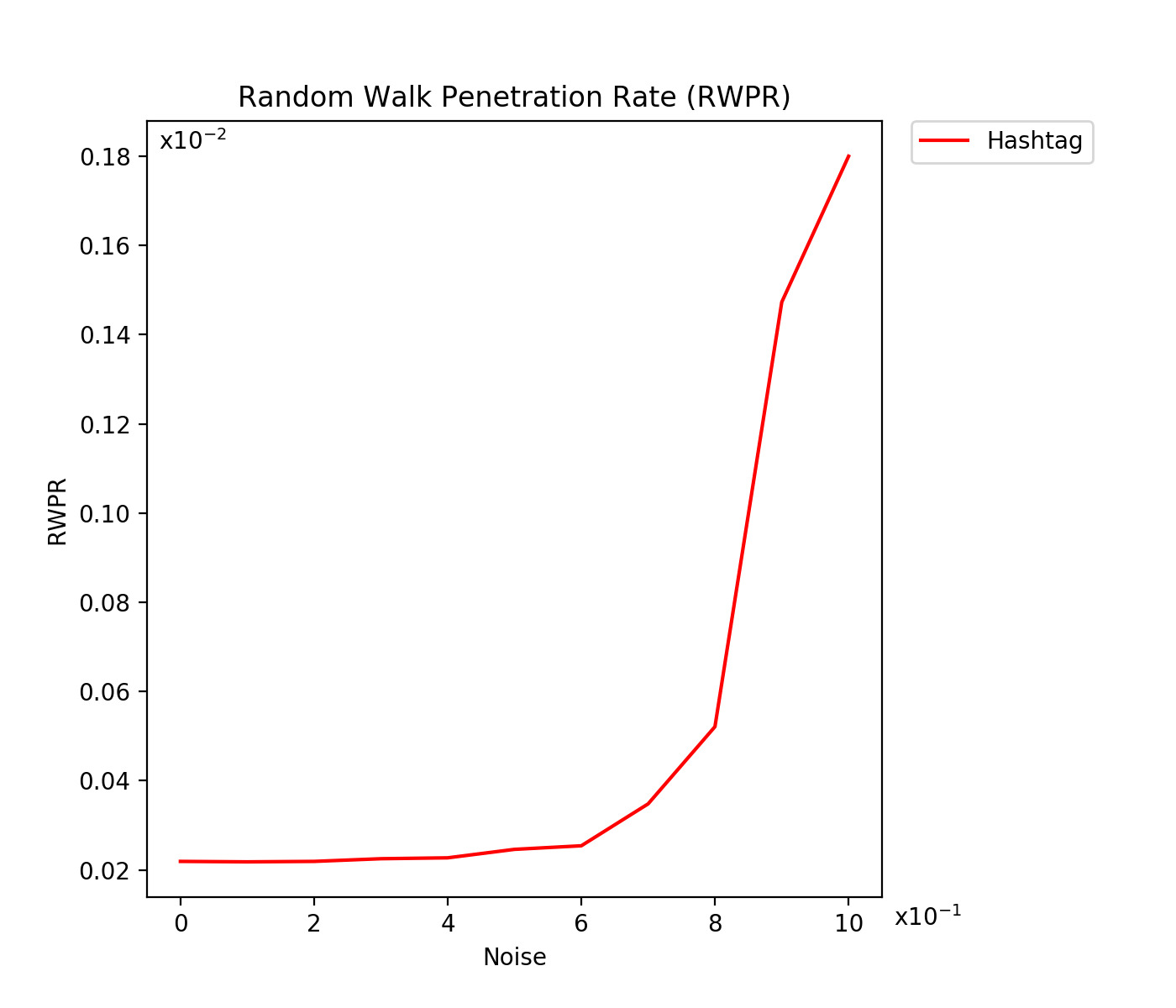}
		\caption{RWPR of hashtag network.}
		\label{Figure_7}
	\end{subfigure}
	\begin{subfigure}[t]{0.3\textwidth}
		\includegraphics[width=\textwidth]{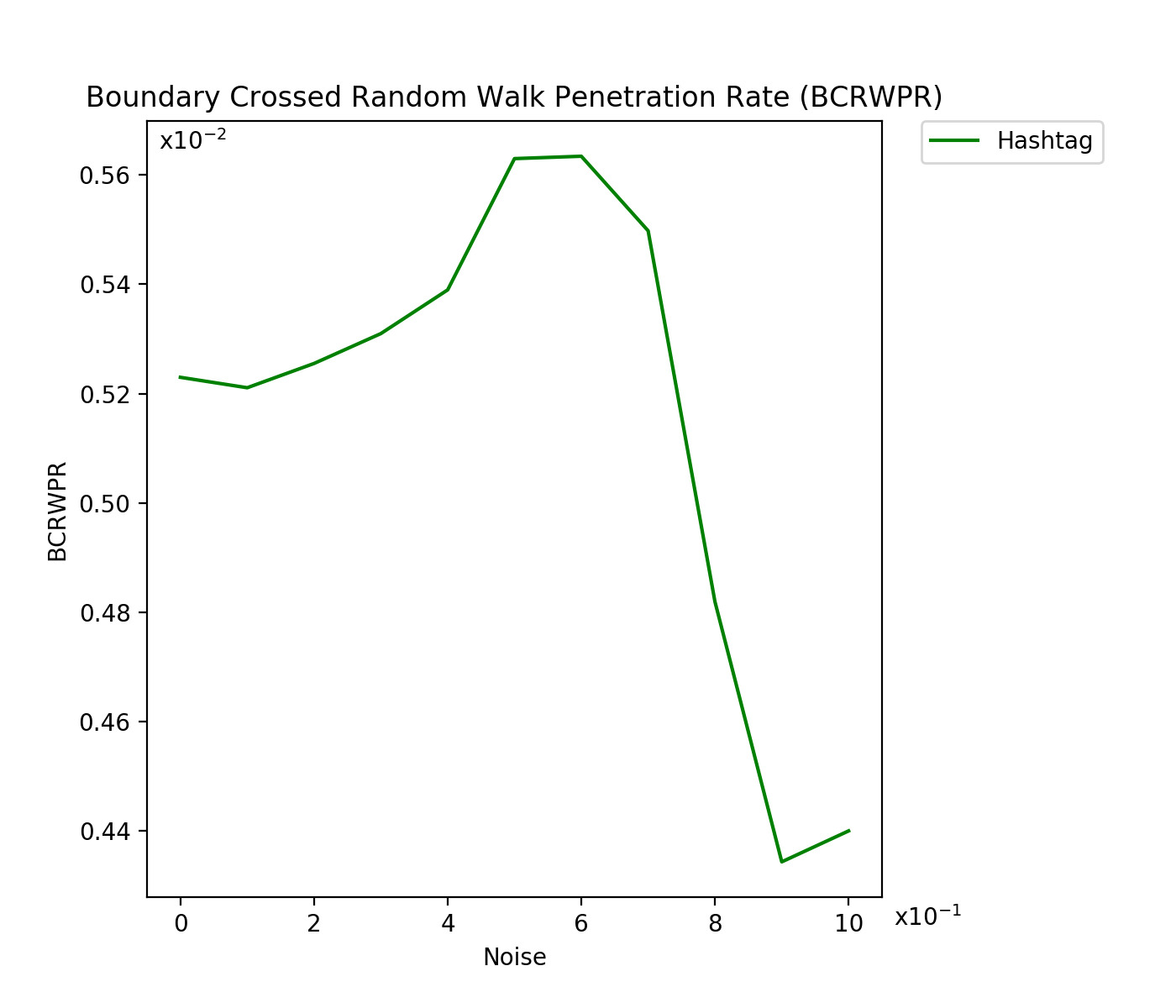}
		\caption{BCRPR of hashtag network.}
		\label{Figure_8}
	\end{subfigure}
	\begin{subfigure}[t]{0.3\textwidth}
		\includegraphics[width=\textwidth]{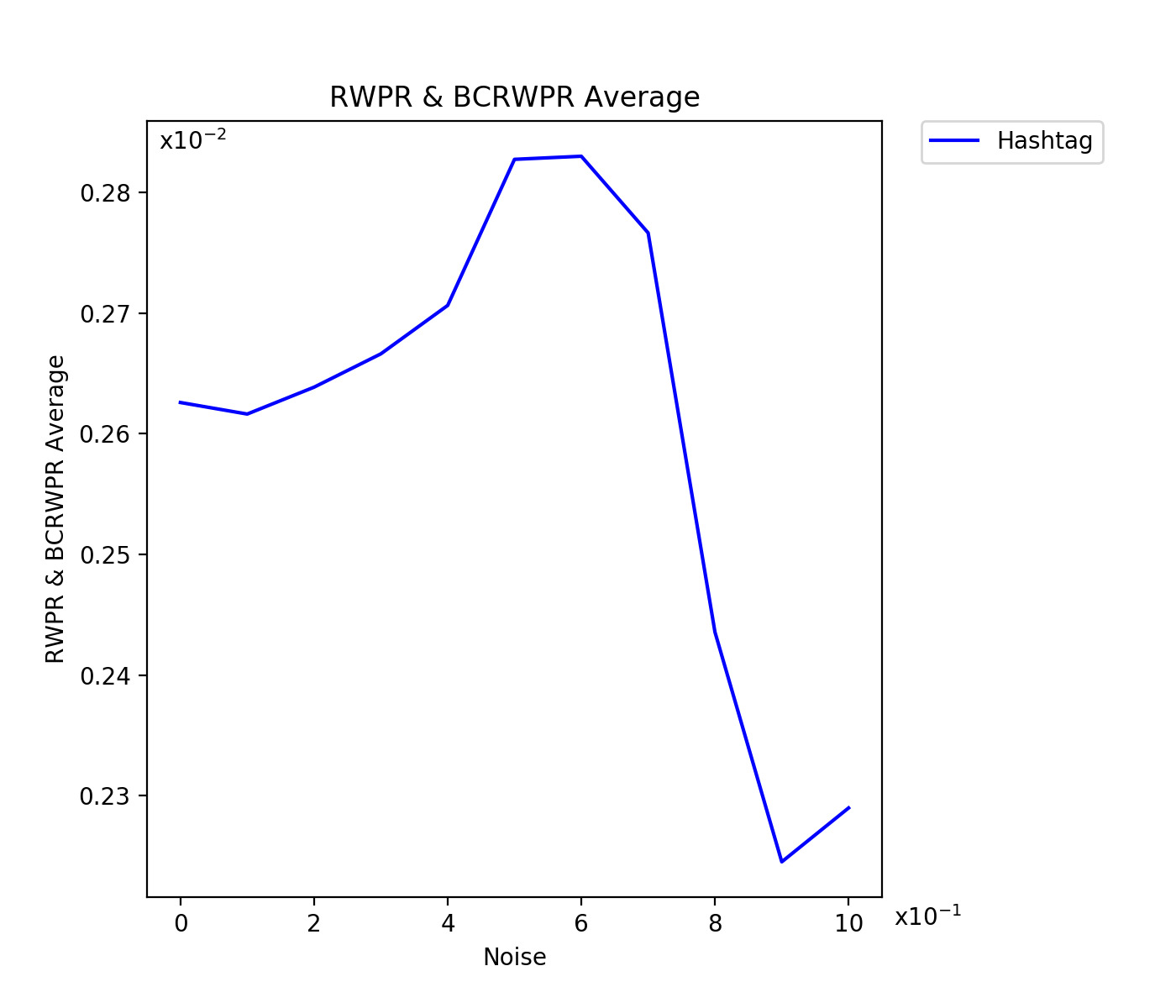}
		\caption{Average of hashtag network.}
		\label{Figure_9}
	\end{subfigure}
	\caption{Graph of BRW behavior with different levels of noise for hashtag graph. These graphs indicate that RWPR, unlike BCRPR and Average of hashtag network, reflect the inject the nature of inject noise correctly.}
	\label{Figure_hashtag}
\end{figure*}

\begin{figure*}
	\centering
	\begin{subfigure}[t]{0.4\textwidth}
		\includegraphics[width=\textwidth]{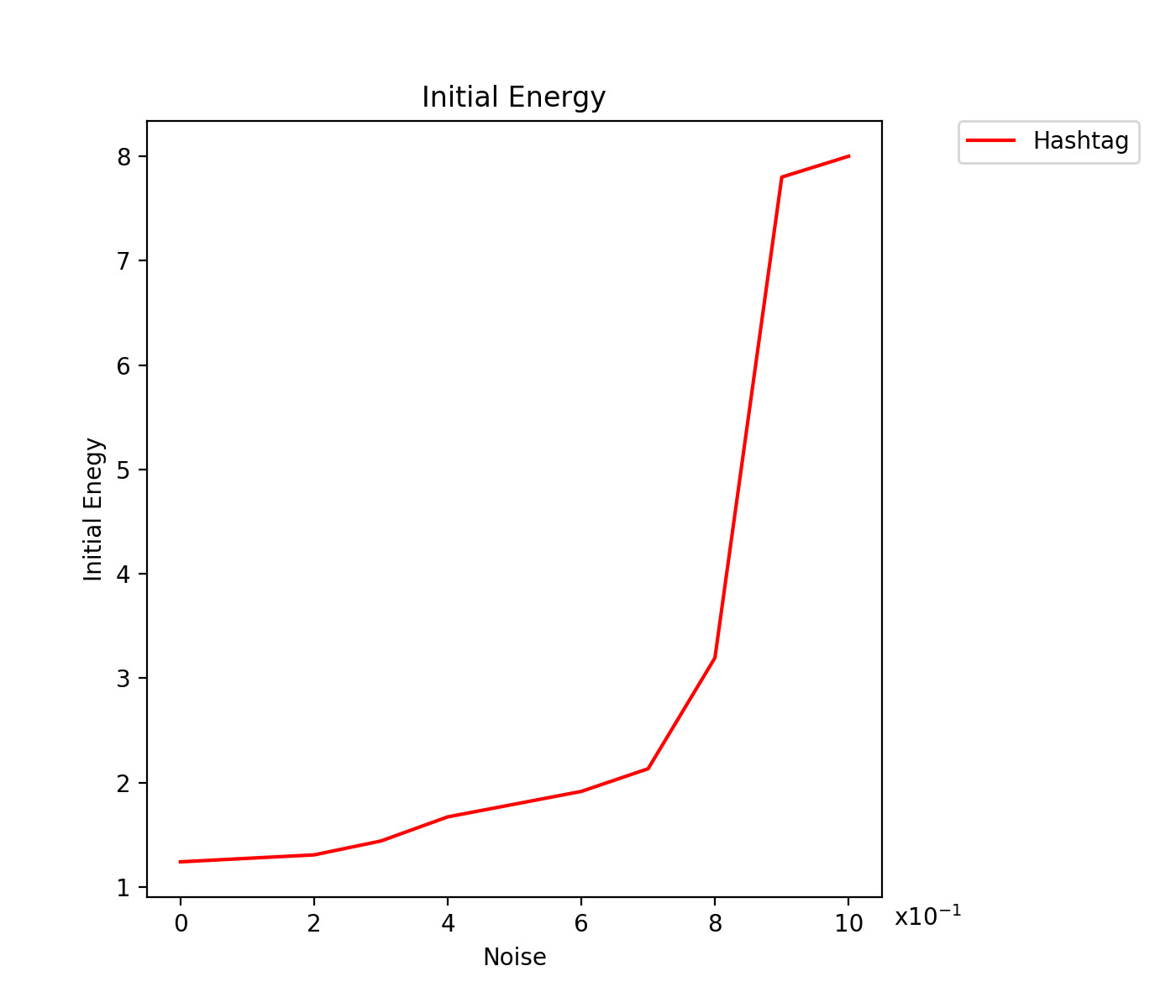}
		\caption{Average initial energy of hashtag network for different levels of noise. }
		\label{Start_3}
	\end{subfigure}
	\begin{subfigure}[t]{0.4\textwidth}
		\includegraphics[width=\textwidth]{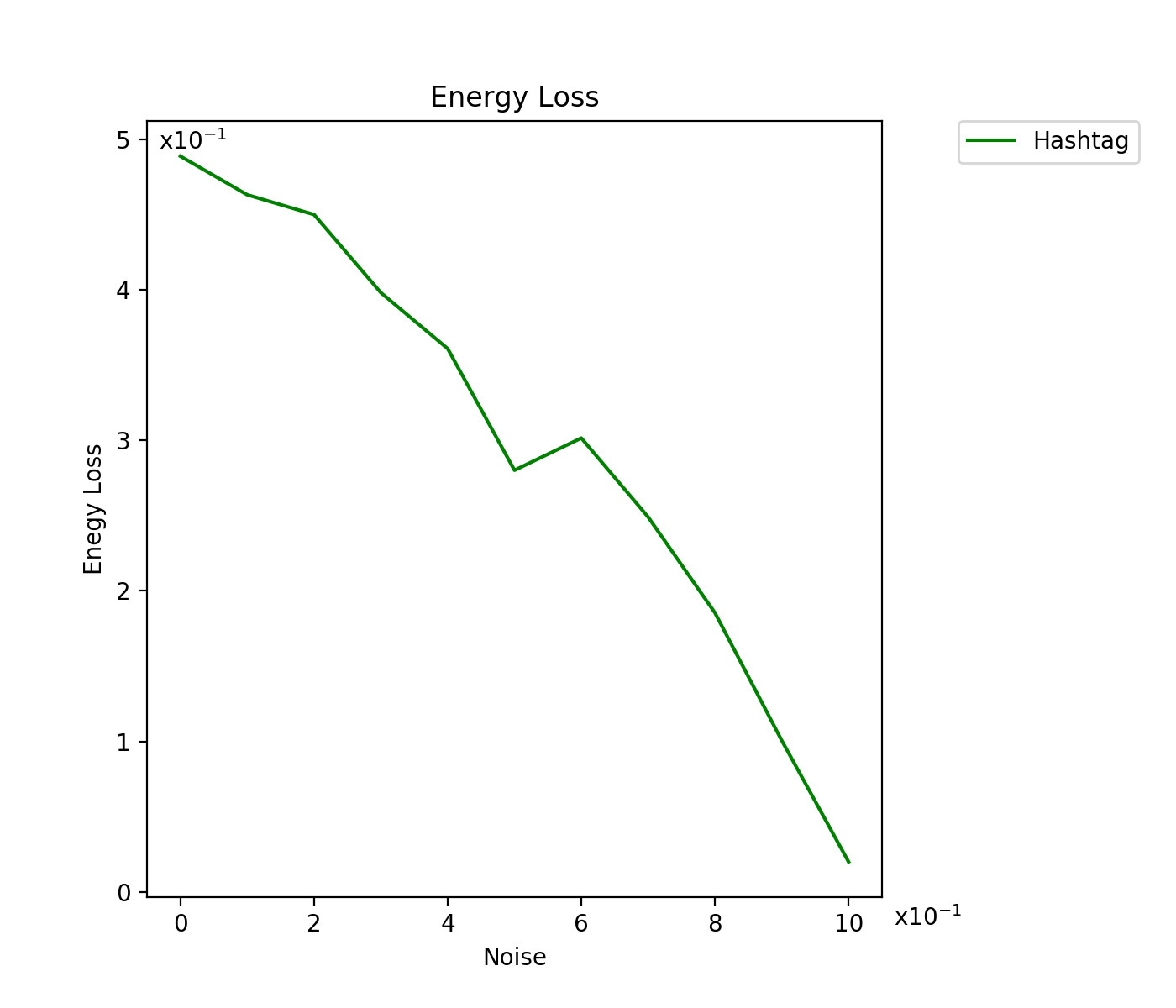}
		\caption{Average energy loss of hashtag network for different level of noise.}
		\label{loss_2}
	\end{subfigure}
	\caption{Average initial energy and energy loss in hashtag network. As it is clear, when we inject noise to hashtag network the average initial energy increase significantly, and the average energy loss decrease.}
	\label{intital_hashtag}
\end{figure*}

The next considered network is hashtag network in which the only features for each node are the reduced vector, by PCA, of hashtag usage. Fig. \ref{Figure_hashtag} shows the behavior of BRW for hashtag network. Fig. \ref{Figure_7} indicates the RWPR. The graph increases slowly in the first steps and suddenly rises significantly at bigger noises. Because it is not evident, we also demonstrate the exact number of Fig \ref{Figure_7} graph in Table \ref{tab5}. Fig. \ref{Figure_8} illustrates that the BCRPR depends on network structure and not other kinds of features, as described before. Fig. \ref{intital_hashtag} shows that as we expect the initial energy of nodes increases when their hashtags become similar, at the same time, the energy loss of nodes decreases, which is indicative of nodes' gentility in passing information from one side of network to another side.

\begin{table}{t}
	\centering
	\caption{Table of RWPR }
	\label{tab5}
	\begin{tabular}[\columnwidth]{|c c|}
		\hline
		Noise &PNC \\
		\hline
		0.0 & 0.002187\\
		\hline
		0.1 & 0.002181  \\
		\hline
		0.2 & 0.002191  \\
		\hline
		0.3 & 0.002247 \\
		\hline
		0.4 & 0.002272 \\
		\hline
		0.5 &0.002460 \\
		\hline
		0.6 & 0.0025403 \\
		\hline
		0.7 & 0.003483 \\
		\hline
		0.8 & 0.005205 \\
		\hline
		0.9 & 0.014729 \\
		\hline
		1.0 & 0.018345\\
		\hline
	\end{tabular}
	
\end{table}

\begin{figure*}
	\centering
	\begin{subfigure}[t]{0.3\textwidth}
		\includegraphics[width=\textwidth]{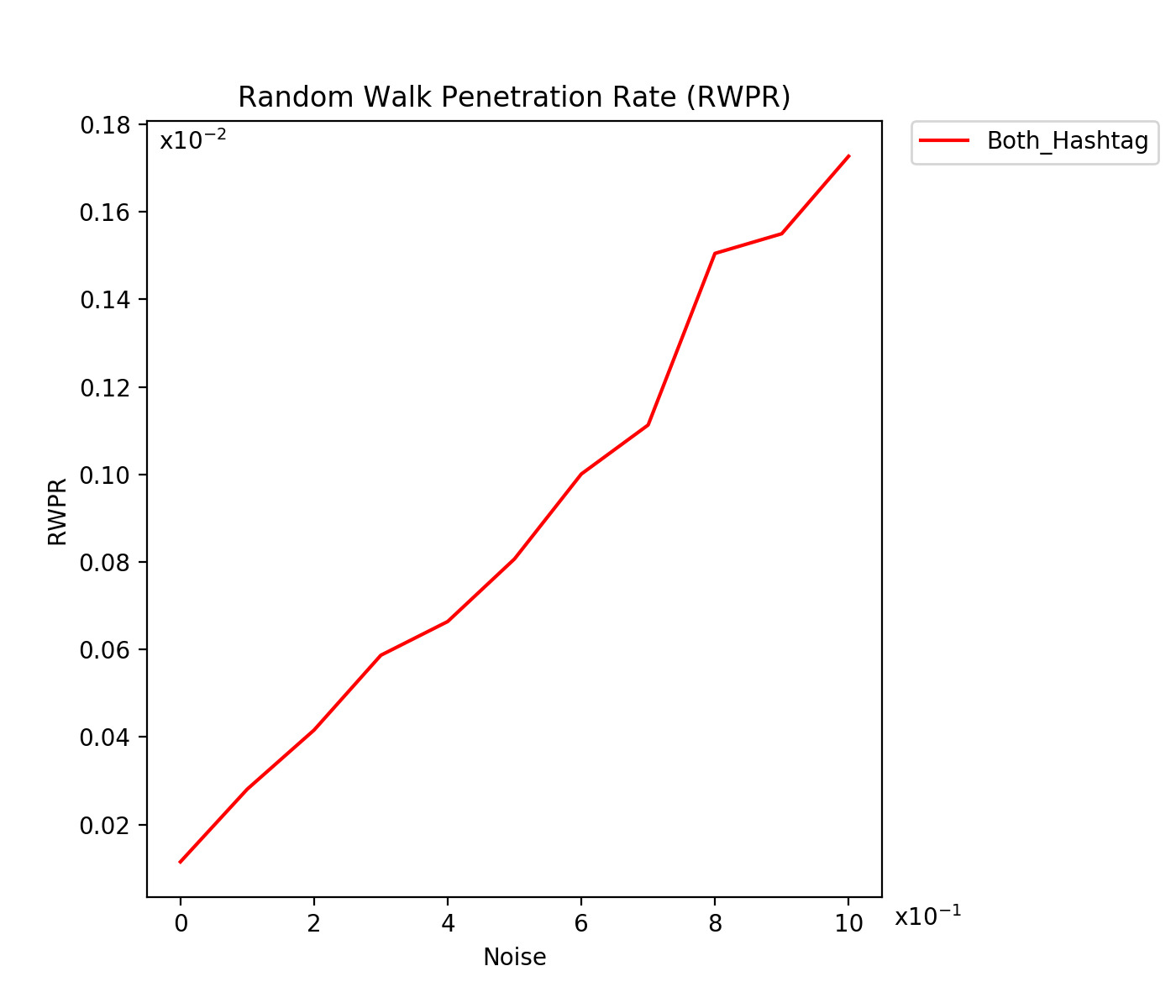}
		\caption{RWPR of both\_hashtag network.}
		\label{Figure_10}
	\end{subfigure}
	\begin{subfigure}[t]{0.3\textwidth}
		\includegraphics[width=\textwidth]{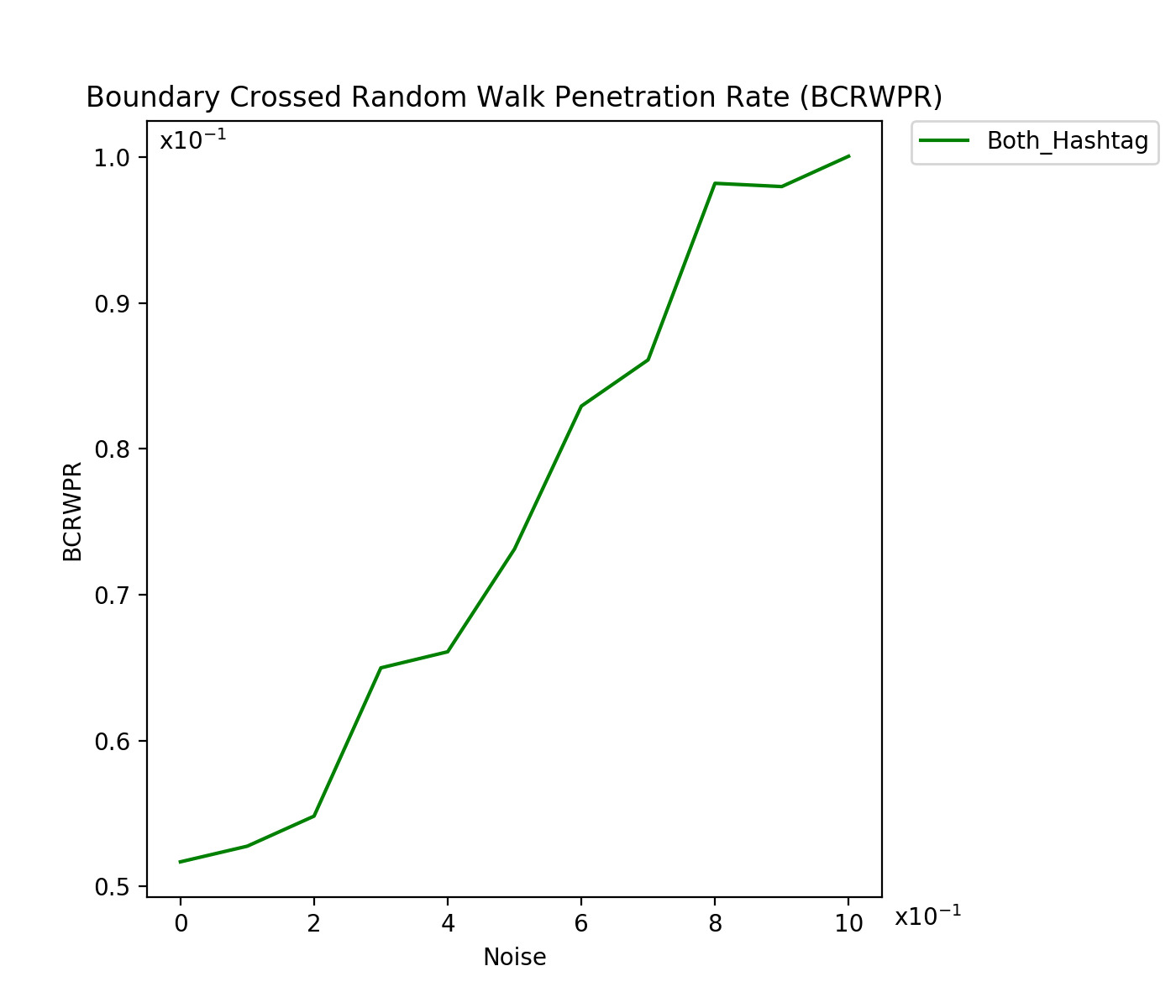}
		\caption{BCRPR of both\_hashtag network.}
		\label{Figure_11}
	\end{subfigure}
	\begin{subfigure}[t]{0.3\textwidth}
		\includegraphics[width=\textwidth]{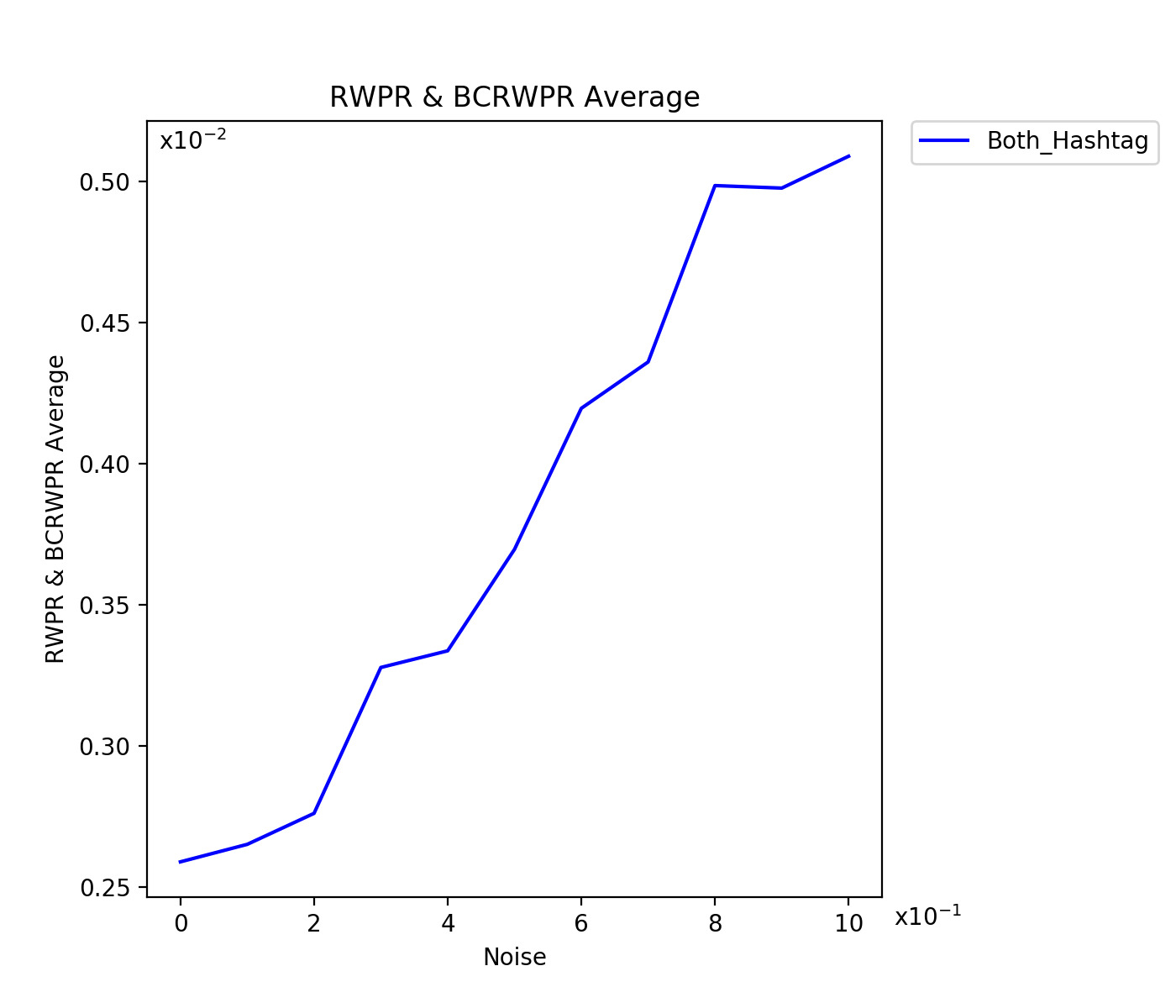}
		\caption{average of both\_hashtag network.}
		\label{Figure_12}
	\end{subfigure}
	\caption{Graph of BRW behavior with different levels of noise for both\_hashtag network. The graphs show that when we inject noise to the networks, as we expect, the controversy level of networks decrease. This is true for all RWPR, BCRPR, and average of both\_hashtag networks.}
	\label{Figure_both}
\end{figure*}

\begin{figure*}
	\centering
	\begin{subfigure}[t]{0.4\textwidth}
		\includegraphics[width=\textwidth]{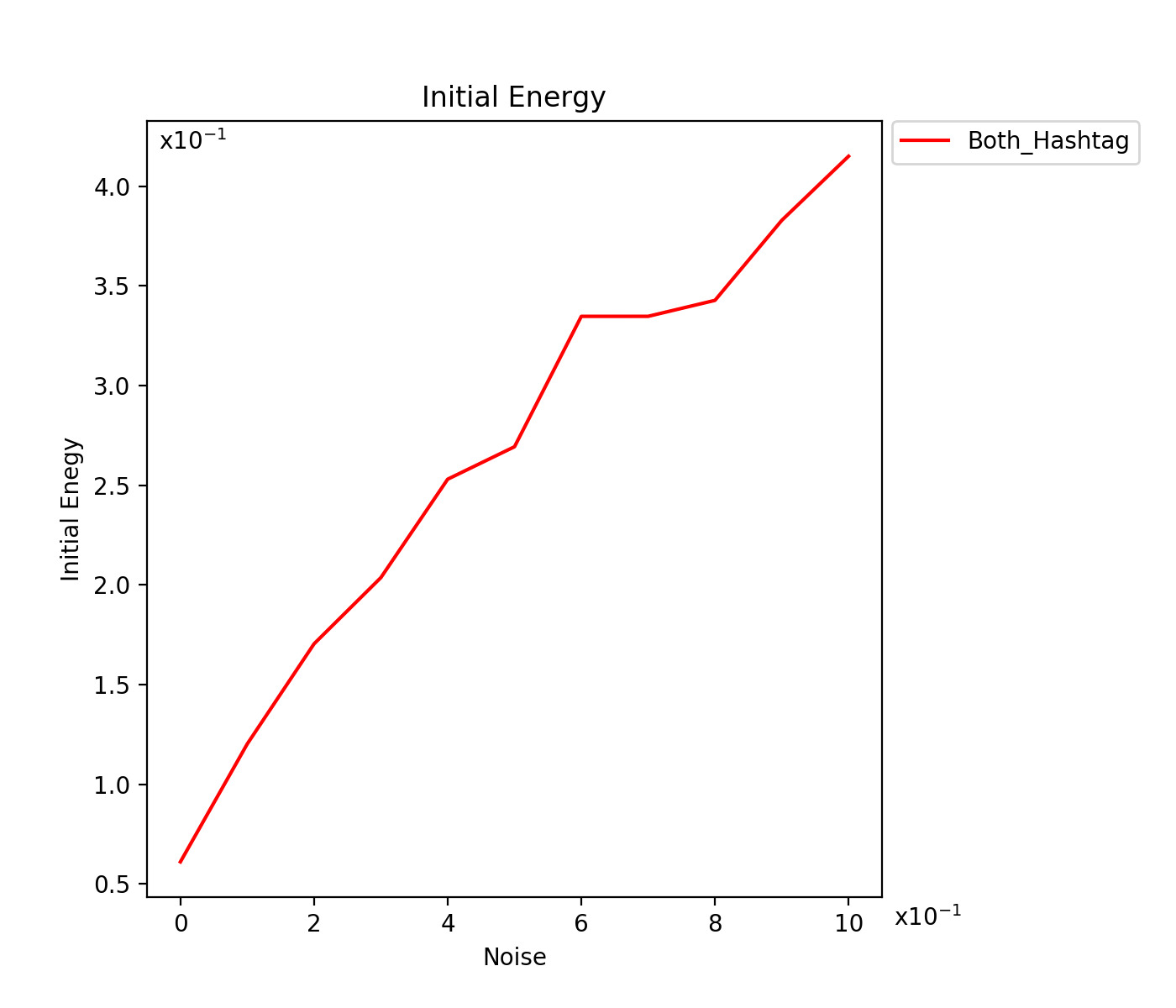}
		\caption{Average initial energy of both\_hashtag networks for different levels of noise.}
		\label{Start_4}
	\end{subfigure}
	\begin{subfigure}[t]{0.4\textwidth}
		\includegraphics[width=\textwidth]{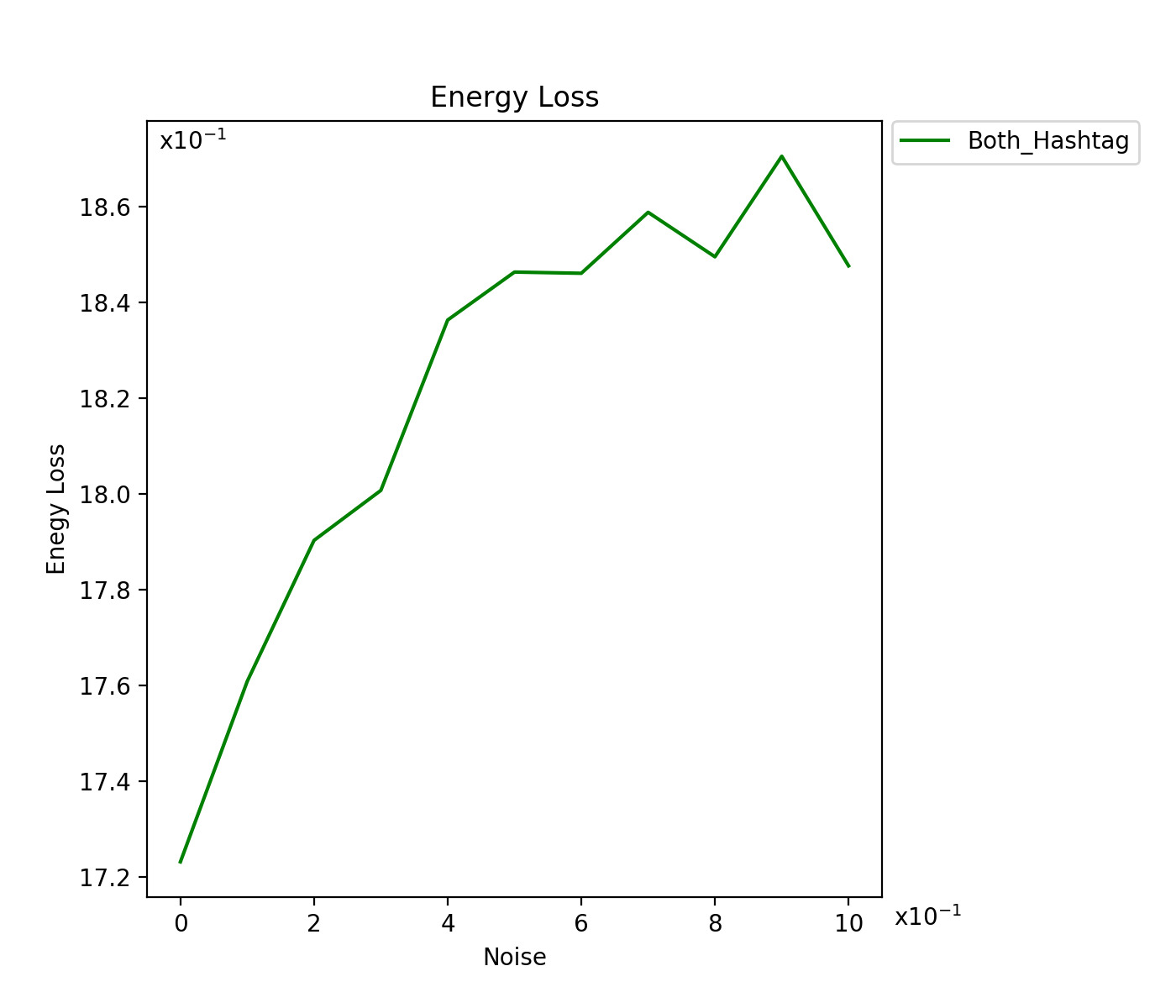}
		\caption{Average energy loss of both\_hashtag network for different level of noise.}
		\label{loss_3}
	\end{subfigure}
	\caption{Average initial energy and energy loss in both\_hashtag networks. This graphs indicate significant increase in the average initial energy compared to smooth increase in the average energy loss.}
	\label{intital_both}
\end{figure*}

Now we consider the behavior of the network when both structural (Node2vec) and content information (hashtags) are taken into account (we call this network as both\_hashtag since we do not have noise on hashtag features to distinguish it with the next subsection's networks in which we have noise on both structure and content information). In other words, in this special case, we track behavior of network when both network and content behavior exist, but we have noise only on structural features. Similar to Node2vec network, in this case, also, RWPR, BCRPR, and consequently average metric increase when we randomly add edges to push well-separated communities toward each other \ref{Figure_both}. The initial energy and energy loss simulate the behavior of Node2vec network, as well, Fig. \ref{intital_both}.

\begin{figure*}
	\centering
	\begin{subfigure}[t]{0.3\textwidth}
		\includegraphics[width=\textwidth]{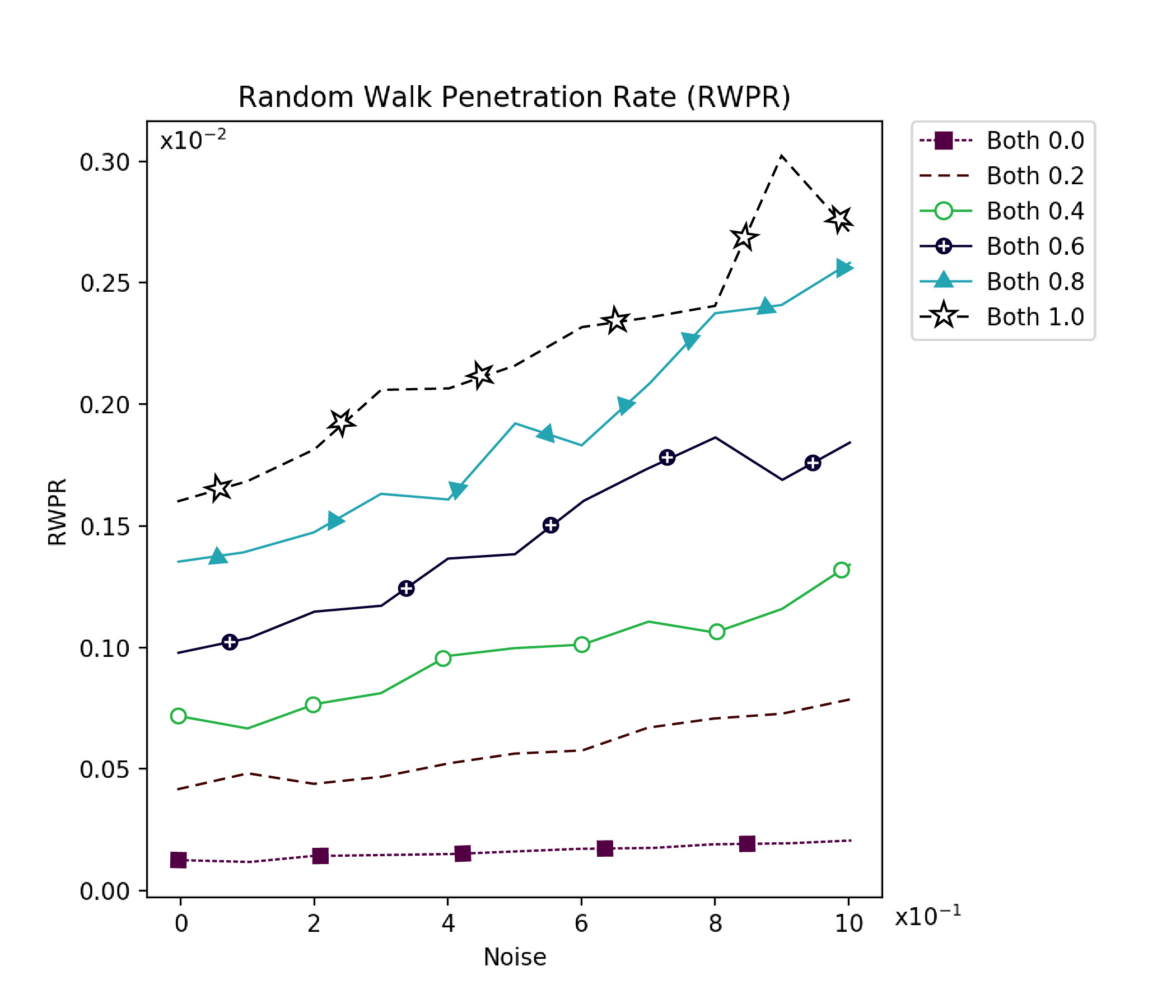}
		\caption{RWPR of both network.}
		\label{Figure_1}
	\end{subfigure}
	\begin{subfigure}[t]{0.3\textwidth}
		\includegraphics[width=\textwidth]{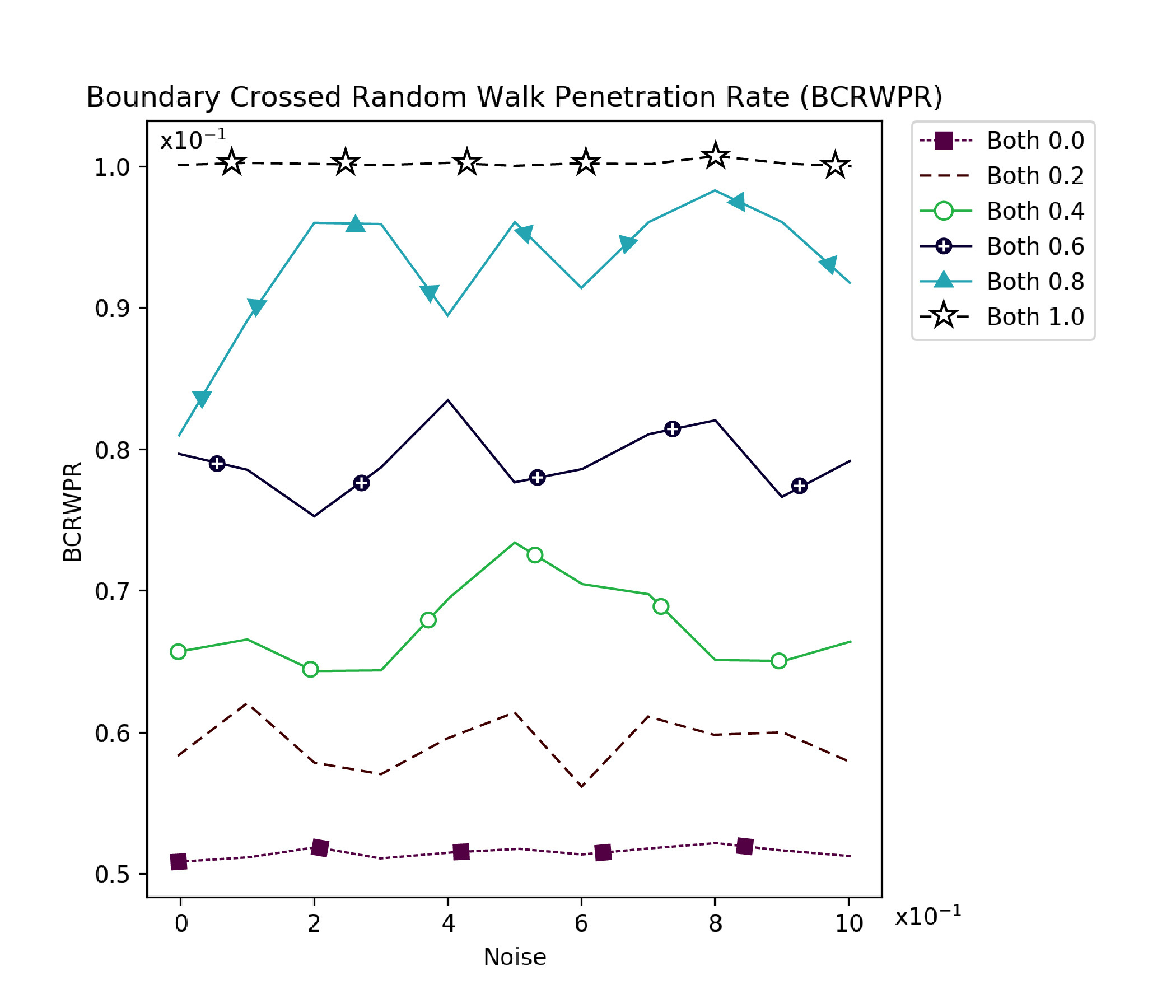}
		\caption{BCRPR of both network.}
		\label{Figure_2}
	\end{subfigure}
	\begin{subfigure}[t]{0.3\textwidth}
		\includegraphics[width=\textwidth]{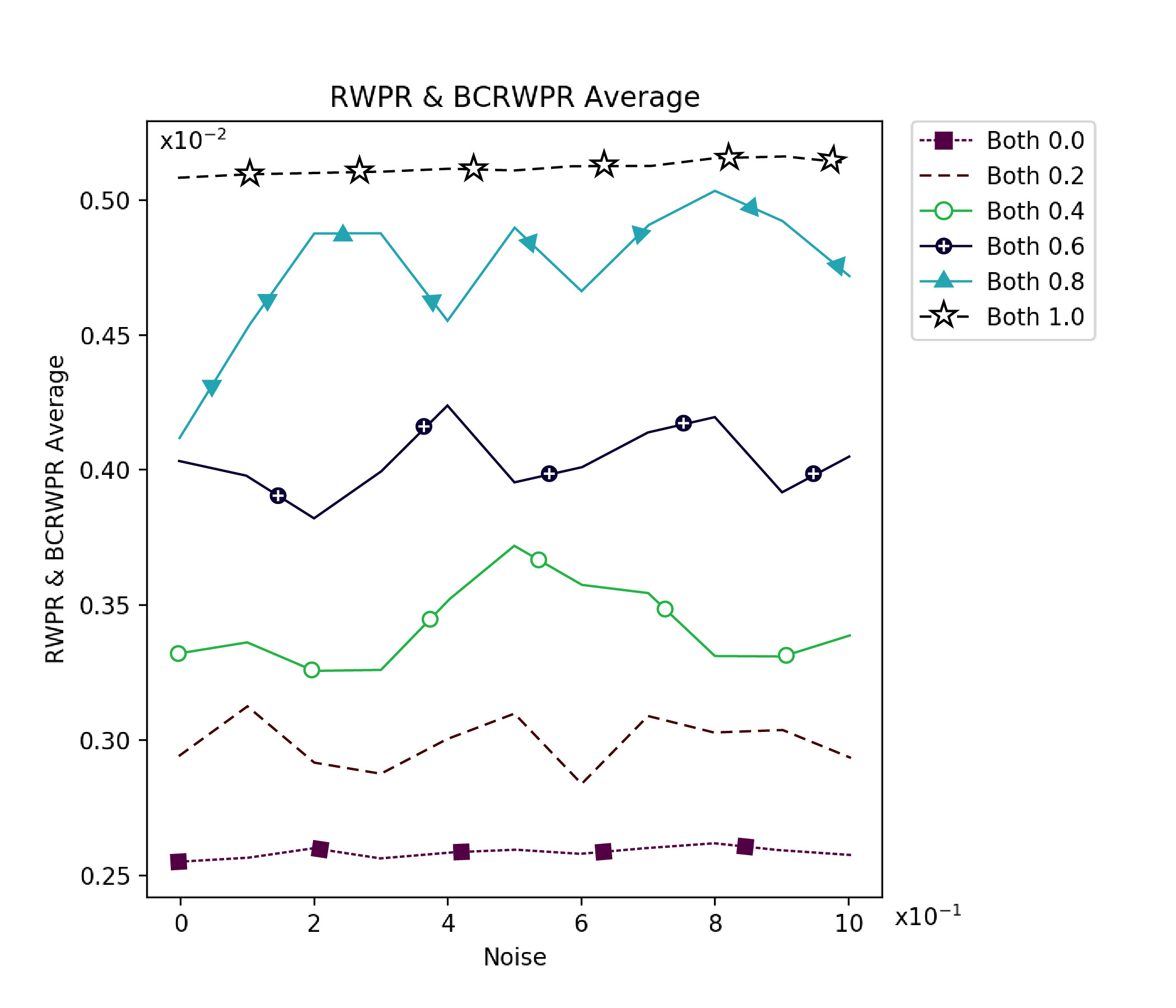}
		\caption{average of both network.}
		\label{Figure_3}
	\end{subfigure}
	\caption{Graph of BRW behavior with different levels of noise for \textit{both} network. This graphs show that RWPR reflect injected noise effect to the network correctly.}
	\label{Figure_both_both}
\end{figure*}

\begin{figure*}
	\centering
	\begin{subfigure}[t]{0.4\textwidth}
		\includegraphics[width=\textwidth]{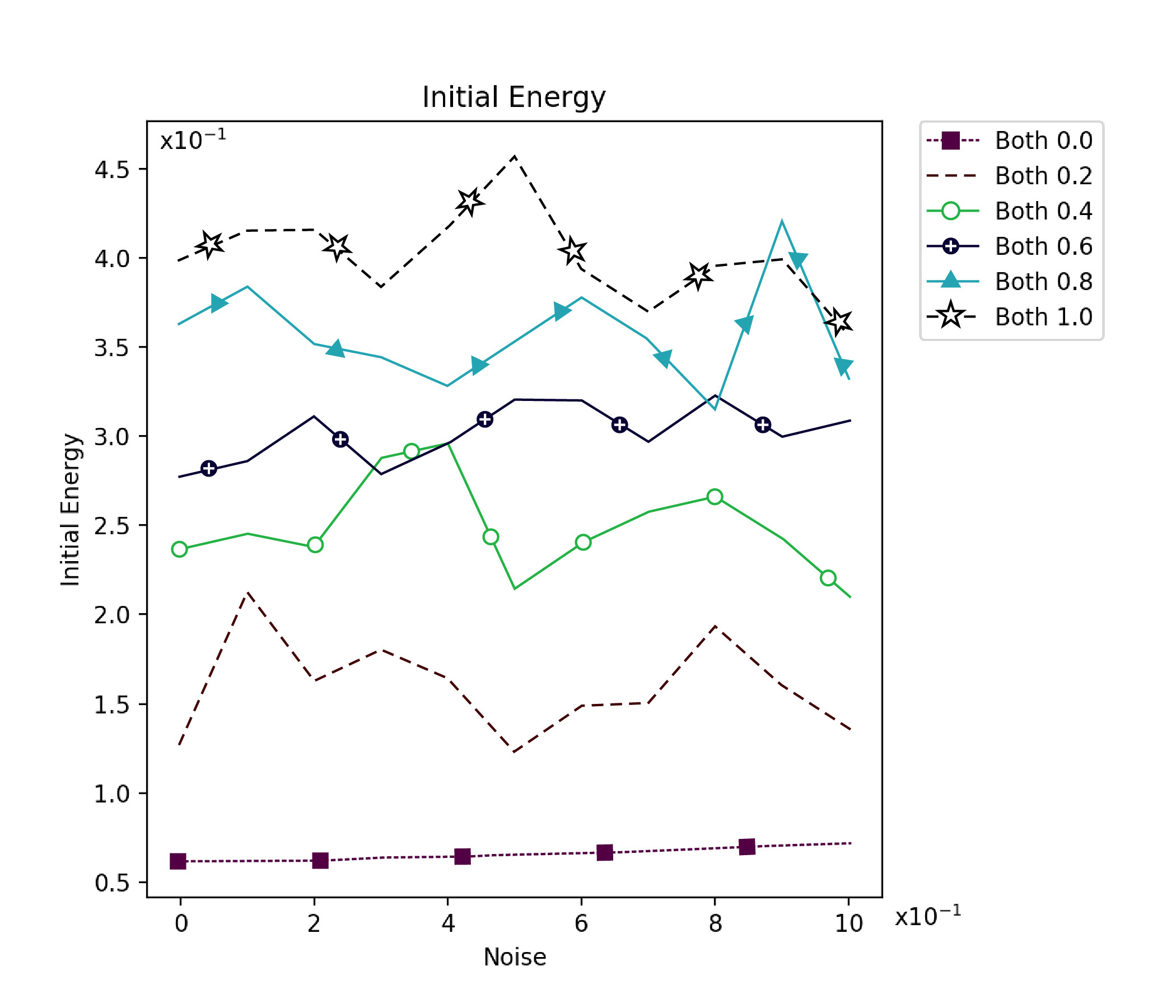}
		\caption{Average initial energy of both network for different level of noise.}
		\label{Start_Energy}
	\end{subfigure}\hspace{15mm}
	\begin{subfigure}[t]{0.4\textwidth}
		\includegraphics[width=\textwidth]{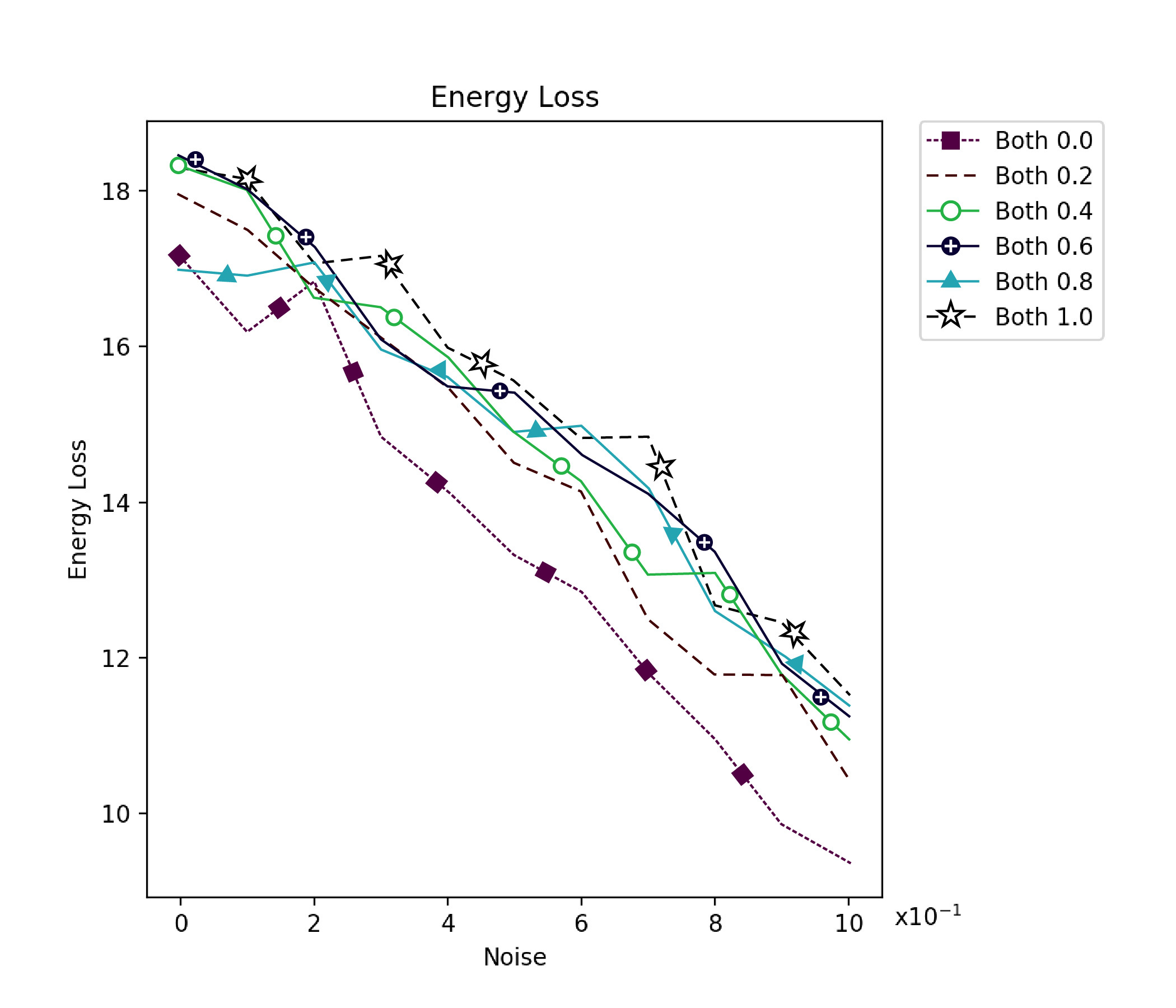}
		\caption{Average energy loss of both network for different level of noise.}
		\label{Loss_Energy_1}
	\end{subfigure}
	\caption{Average initial energy and energy loss in both network. These graphs show the effect of noise on the average initial energy and average energy loss when we inject the noise to both the hashtags and network.}
	\label{intital_both_both}
\end{figure*}

Finally, we evaluate the changes in the behavior of BRW when we impose noise on both structural and content information. For this case, we add five levels of structural noise to the networks(0.2, 0.4, 0.6, 0.8, 1.0), then, for each case, we calculate their controversy level by adding noise to content features. Fig \ref{Figure_both_both} indicates the behavior of five network produced by adding noise to their structures. Additionally, x axis shows the imposed noise on the content feature of networks. By adding the noise; therefore, making the nodes similar to each other increase RWPR, but BCRPR remain almost the same. Although initial energy increase in bigger structural noise, there is no sign of influence imposed by content noise in this case, Fig \ref{Start_Energy}. Maybe this originates from sparseness of content date. Energy loss for the network decreases, as expected \ref{Loss_Energy_1}.

\subsection{Iranian tweets in three months}
For comparing the changes of political and non-political topics controversy level in Persian tweets we made 5, versions of networks. Then, the results for Node2vec, Hashtag, Both, in addition to RW is presented for comparing the results of our algorithm with RW as the baseline method, which outperforms other controversy measuring methods \cite{guerra2013measure}.
\subsubsection{Data Set}
First of all, we extracted all the tweets which contained at least one of the hashtags which happened more than150 times among all the tweets, collected from November first to January 29th. The established network on this data set is the network which we call it \textbf{All} in the table. All\_politic is the network which we made this network by extracting all political hashtags with more than 150 occurrences among the all tweets (pay attention that in the section \ref{subsub}, we considered all the tweets with political hashtags regardless of the hashtags frequency). Finally, we separated \textbf{All} network with respect to the tweet date to Nov., Dec., and Jan. (Table \ref{tab7}). \\

In the past few years, Iran has encountered several protests regarding the political and political economic issues. Therefore, we have to consider the political issues much more controversial, as they end up with street protestations as the maximum level of controversy compared to daily life events.\\

 Due to bipolarity level in Iran and the small number of boundary nodes, most of the random walks could not end up in another community, if any. For having more realistic perspective in this section, we multiplied the initial energy to 200, this is the case for all the runs; the other variables of configuration remained the same as before.\\

\subsubsection{Results}

Table \ref{tab8} shows the RWPR and RW of the networks. As expected, the network containing all topics is less controversial, and this is evident in bigger RWPR for \textbf{All} network compared to All\_politic network. Nov. has bigger RWPR in comparison to other months, and this is because of the protests which occurred in this month as reaction to increase in petrol price. Jan. has the smallest RWPR due to the death of Iran's popular General which followed by one of the biggest and glorious funeral ceremony in Iran history. The results match with the real atmosphere in Iran, and this is true for hashtag, and both but not Node2vec. The result of Node2vec is precisely similar to RW as they both take into account only structural information. \\

Table \ref{tab8} indicates that RW correctly demonstrates \textbf{All} and All\_politic networks relations. However, it is unable to show Nov. and Dec. relations, correctly. We know that due to nationwide protestations Twitter atmosphere beside the social atmosphere was controversial in Nov. Whereas, in Dec. Iran had only occasional protestations. Thus, the Twitter atmosphere beside the social environment was less controversial. But RW, similar to Node2vec is not able to reflect this relation in their results.\\

In fact, the difference between Nov. and Dec. indicates the changes in controversy level of the frequently discussed topics in Persian tweets. Therefore, inability of RW and Node2vec reflects importance of other sources of information for improving precision of the algorithms. The correct order of controversy level, detected by our algorithm, not only demonstrates that our algorithm out performs RW, but also indicates that more information should be taken into account for better results.\\

These results prove that, in some cases, using only structural data, we are not capable of evaluating the controversy level of the social network environment. In turn, by combining textual and profile, we can overcome this shortcoming. 
\begin{table*}
	\centering
	\caption{Information of Persian tweets networks.}
	\label{tab7}
	\begin{tabular}{|l r r l|}
		\hline
		Name &  \#Node & \#Edge & Description and collection period \\
		\hline
		All &  992 & 3815 & All tweets from Nov 1 to Jan 29\\
		\hline
		All\_politic  & 467 & 1278 & All political tweets from Nov 1 to Jan 29\\
		\hline
		Nov  & 718 & 2706 & All tweets from Nov 1-30 \\
		\hline
		Dec  & 216 & 577 & All tweets from Dec 1-31\\
		\hline
		Jan  & 211 & 537 & All tweets from Jan 1-29\\
		\hline
	\end{tabular}
	
\end{table*}

\begin{table}
	\centering
	\caption{These table indicate controversy level of networks, measured by RWPR and RW. These networks are the networks created using all Persian tweets, all political Persian tweets, Persian tweets in Nov. Dec., and Jan. Iran's society experienced different level of bipolarity in this period of time.}
	\label{tab8}
	\begin{tabular}{|l r r r r l|}
		\hline
		Name &  Node2vec & Hashtag & Both & RW & Description\\
		\hline
		All &  0.02165 & 0.0044 & 0.0034 & 0.582 & All Persian events\\
		\hline
		All\_politic & 0.0299 & 0.0013 &  0.0013 & 0.74 & Political events\\
		\hline
		Nov  & 0.03083 & 0.0019 & 0.0014 & 0.68 & Nationwide protestations\\
		\hline
		Dec  & 0.0233 & 0.0023 & 0.0021 & 0.76 & Occasional protestations\\
		\hline
		Jan  & 0.0530 & 0.0025& 0.0024 & 0.65 & Iran's general Funeral\\
		\hline
	\end{tabular}
	
\end{table}

\section{Conclusion}
\label{sec_conclud}
This paper propose a flexible framework for taking into account the non-structural features of network provided in social media context. Previous research focused on structural information, while they neglected precious information source provided by users in social networks. Our implementation of the proposed frame work (BRW) is a biased random walker, which starts with an initial energy proportionate to the position of starting node and loss its energy in each step proportionate to the energy loss of each node on the path. The level to which a random walk reaches in the contradicting community (the community which does not belong to) is its ability in penetration, in other words, it is being heard. We call this controversy of a topic.\\

It seems that community detection based models for assessing controversial topics have the inherent defects depending on the basic community detection algorithm. For future research, we are going to release controversy detection measures from community detection algorithms to have more realistic perspective to controversial topic issue.

‎


\bibliographystyle{cas-model2-names}

\bibliography{cas-refs}


\end{document}